\documentclass[prx,preprint,preprintnumbers]{revtex4}
\usepackage{graphicx} % Include figure files; subfigure
\usepackage{dcolumn} % Align table columns on decimal point
\usepackage{bm} % bold math
\usepackage{amsmath}

\usepackage[dvips]{color}

\usepackage{epsfig}
\newcommand{\nn}{\nonumber}

\begin{document}
\title{Coupled-mode theory for electromagnetic pulse propagation in dispersive media undergoing a spatio-temporal perturbation - exact derivation, numerical validation and peculiar wave mixing}

\author{Y. Sivan, S. Rozenberg, A. Halstuch} % , Shlomo Pinhas}
\affiliation{Faculty of Engineering Sciences, Ben-Gurion University of the Negev, Israel} % Unit of Electro-optics
\email{sivanyon@bgu.ac.il}

% PRX!!! 1700$, IF 8, 20,000 words, worst case, goes to PRA.

% post sending to Alon and Gadi..

\begin{abstract}
   We present an extension of the canonical coupled mode theory of electromagnetic waves to the case of pulses and spatio-temporal perturbations in complex media. Unlike previous attempts to derive such a model, our approach involves no approximation, and does not impose any restriction on the spatio-temporal profile. Moreover, the effect of modal dispersion on mode evolution and on the coupling to other modes is fully taken into account. Thus, our approach can yield any required accuracy by retaining as many terms in the expansion as needed. It also avoids various artifacts of previous derivations by introducing the correct form of the solution. We then provide what is, to the best of our knowledge, the first ever validation of the coupled mode equations with exact numerical simulations, and demonstrate the wide range of possibilities enabled by spatio-temporal perturbations of pulses, including pulse shortening or broadening or more complex shaping. Our formulation is valid across the electromagnetic spectrum, and can be applied directly also to other wave systems.
\end{abstract}


%42.65.Sf	Dynamics of nonlinear optical systems; optical instabilities, optical chaos and complexity, and optical spatio-temporal dynamics

% 42.65.Wi	Nonlinear waveguides

% 42.65.-k	Nonlinear optics

%42.79.Dj	Gratings
%
%42.79.Gn	Optical waveguides and couplers
%

\maketitle

\section{Introduction}

Decades of research have given us a deep understanding of electromagnetic wave propagation in complex (structured) media. This includes, specifically, plane wave, beam and pulse propagation in media having any length scale of structuring, from slowly graded index media (GRIN) where the structure is characterized by a length scale of many wavelengths, through photonic crystals to metamaterials, where the structure can have deep subwavelength features. Analytical, semi-analytical and numerical tools are widely available for such media along with a wide range of applications.

On the other hand, our understanding of wave propagation in media whose electromagnetic properties vary in space {\em and time} is less developed. Remarkably, this situation occurs for all {\em coherent} optical nonlinearities, described via a nonlinear susceptibility~\cite{Boyd-book}, in which case the associated nonlinear polarization serves as a spatio-temporal perturbation (or source). This also occurs for the interaction of an electromagnetic pulse with a medium having an {\em incoherent} optical nonlinearity such as for free carrier excitation~\cite{Kost-book,Lipson_review,Euser_Thesis,Notomi_review} or thermal nonlinearities~\cite{Boyd-book,Stoll_review}. The electromagnetic properties vary in time also for moving media in general~\cite{Sommerfeld-book}, and specifically, in the context of special relativity (see~\cite{Philbin-Leonhardt-event-horizon,Biancalana_1D_layered_media_n(t)} and references therein); they also undergo peculiar modifications for accelerating perturbations~\cite{Bahabad_accelerating}.

Spatio-temporal perturbations can originate from a perturbation induced by an incoming wave, as for harmonic generation~\cite{Boyd-book} or for pulses propagating in Kerr media (e.g., for self-focusing or self phase modulation, see~\cite{Agrawal-book}). They can also originate from the interaction of several waves of either comparable intensities, as in many nonlinear wave mixing configurations, or of very different intensities as, e.g., in pump-probe (or cross phase modulation~\cite{Agrawal-book}) configurations.

Temporal variation of the electromagnetic properties can enable various applications such as switching in communication systems~\cite{Tajima-exp,Wada_2004,Notomi_review,Lipson_review,Nielsen-Mork,Lincoln-lab} (especially of pulses), pulse broadening and compression~\cite{Biancalana_1D_layered_media_n(t)}, dynamic control over quantum states in atomic systems~\cite{Harris-Yamamoto-switching,Matsko,EIT_review,Lukin_switching,Gauthier-Rubidium-switching,Gaeta_rubidium_switching} or in complex nanostructures~\cite{FC_general_1,Gerard_review,Gerard_spEm_control,Vos_PRB_2002,Wehrspohn_PRB_2002,Vuckovic_PhC_switching,Twamley_cavities_switching,Henri_ZAP_theory,Henri_micropillar_switching}, non-reciprocal propagation~\cite{Fan_interband_isolator,Fan_Lipson_interband_isolator,Russell_isolator,Hadad_Alu_isolator}, time-reversal of optical signals for abberation corrections~\cite{Brener_Fejer_mid-span,Fan-reversal,Fan-reversal-loss,Longhi-reversal,Sivan-Pendry-letter,Sivan-Pendry-article,Sivan-Pendry-HSM}, phase-matching of high-harmonic generation processes~\cite{Bahabad_NP}, control of transmission and coupling~\cite{Fan-CROW-no-dispersion,Rotenberg_OL_2008}, adiabatic wavelength conversion~\cite{Fan-OPN,notomi_dynamic_cavity_PhC_wg,Kuipers_Krauss,Notomi_review} % material science [24-26] of the Armenians..
and more.

% {\color{red}
Wave propagation in uniform media undergoing a purely temporal perturbation was studied in a variety of cases - abrupt (or non-adiabatic/diabatic) perturbations were studied based on the field discontinuities, identified first in~\cite{Morgenthaler}, and then used in several follow up studies, see e.g.~\cite{Felsen_1,Felsen_2,Fante,Mendonca-Guerreiro,Xiao-Agrawal-temporal-boundary,active-Lorentzian,Fink-Cauchy-problem}. Some additional simple~\cite{Agrawal-book} and special~\cite{Shvartsburg} cases have exact solutions, but general smooth time-dependent perturbations require sophisticated asymptotics, see e.g.~\cite{Armenians_smooth_t_perturbation}. Studies of wave propagation in media undergoing a spatio-temporal perturbation were limited to mostly 3 different regimes, which enable some analytical simplicity. Periodic (monochromatic) perturbations (in space and time) arising from an (explicit) coherent nonlinear polarization are treated within the framework of nonlinear wave mixing~\cite{Boyd-book} while phenomenological~\cite{Kaasik-periodic-time-epsilon,Halevi1,Halevi2} or externally driven~\cite{Katko-PRL-2010} time-periodic perturbations (without spatial patterning) are studied by adapting methods used for spatially-periodic media, such as the Floquet-Bloch theorem. On the opposite limit, quite a few studies involved adiabatic perturbations, in which the temporal perturbation is introduced via the refractive index in uniform media, see, e.g.,~\cite{Mendonca-Guerreiro,Xiao-Agrawal-time-transform}, or phenomenologically via a time-variation of the resonance frequency~\cite{Haus-book,Fan-reversal-loss,Fan-reversal,Fan-CROW-no-dispersion,Fan-OPN,Longhi-reversal} or the transfer function~\cite{Kuipers_Krauss} for complex structured media. In these cases, monochromatic wave methods are usually applicable.

% The fundamental aspects of wave propagation in time-varying media are well-understood - field (dis)continuities upon abrupt changes were identified in~\cite{Morgenthaler} and the spectrum dynamics follows (the violation of) the Noether theorem - a perturbation of the optical properties gives rise to a change of the canonically-conjugate properties of the waves in the system, namely, while a change of the refractive index {\em in space} causes a change of the wave {\em momentum}, a change of the optical properties {\em in time} leads to a change of the {\em spectral} content of the wave. This knowledge enabled the

However, the general case of pulse propagation in time-varying {\em complex} media undergoing a perturbation of arbitrary spatial and temporal scales, has been studied only sparsely (see~\cite{Sipe_Salinas_CMT_deep_gratings_OC,deep_gratings_PRE_96,non_uniform_grating_rigorous,Busch_APB,Reed_doppler,Reed_shock,review_nlo_in_PhC,Sivan-Pendry-letter,Sivan-Pendry-article,temporal_CMT_gratings_russians}), partially because of the complexity of the description (usually involving a system of nonlinearly coupled equations and several interacting waves) and because the theoretical tools at our disposal for handling such problems are limited compared with the case of time-independent media.

Analytical solutions for such configurations are rare, and most studies resort to numerical simulations. The most accurate numerical approach for such problems is the Finite Difference Time Domain (FDTD) method~\cite{Taflove-book}. FDTD is applicable to any structure, as long as the nonlinear mechanism is described self-consistently via a nonlinear susceptibility, $\chi^{(j)}$, a formulation which is implemented in most commercial FDTD packages. Other nonlinearities require the coupling of the Maxwell equations to another set of equations, necessitating dedicated coding (see e.g.,~\cite{FDTD_FC_Agrawal_2010}. Either way, since FDTD provides the field distribution in 3 spatial variables and in time, it is a rather heavy computational tool which is thus limited to short to moderate propagation distances/times. Moreover, like any other pure numerical technique, it provides limited intuition with regards to the underlying physical mechanism is dominating the dynamics.

A possible alternative is the simpler (semi-analytic) formulation called Coupled Mode Theory (CMT), in which the field is decomposed as a sum of monochromatic solutions (aka modes). As the modes are usually constructed from coordinate-separated functions, one can obtain a substantially simpler set of equations involving only the longitudinal spatial coordinate, where the amplitudes of the interacting modes are coupled by the perturbation. CMT is of great importance because it isolates the dominant interacting wave components, facilitates solutions for long distance propagation, can account for any perturbation, including phenomenological ones, and most importantly, it is easy to code and does not require extensive computing resources. On the other hand, CMT is useful only when a limited number of modes are interacting in the system. Fortunately, this is the case in the vast majority of configurations under study.

There are several formalisms of CMT for electromagnetic waves, describing a plethora of cases~\cite{Koegelnik,Yariv-Yeh-book,Haus-book,Chuang_CMT,Hardy_CMT,vectorial_CMT,CMT_Fan_Miller}. The overwhelming majority of these formalisms relates to electromagnetic wave propagation in time-{\em in}dependent media undergoing a purely spatial perturbation (see, e.g.,~\cite{Yariv-Yeh-book}) or to a spatio-temporal perturbation induced by a monochromatic wave, see e.g.,~\cite{Winn_Fan_Joann_Ippen_interband,Busch_APB,temporal_CMT_Bahabad}. Moreover, % in the few cases where {\em pulse} propagation was considered,
dispersion was usually taken into account phenomenologically and/or to leading order only~\cite{Busch_APB,temporal_CMT_Hardy,Sivan-Pendry-HSM,Sivan-COPS-switching-TBG}; in many other cases, it is neglected altogether.
%}

Thus, it is agreed that, to date, there is no universal formalism for treating pulse propagation in dispersive media undergoing a spatio-temporal perturbation. A step towards filling this gap was taken by Dana {\em et al.}~\cite{temporal_CMT_Bahabad}, where the standard CMT formulation of time-{\em in}dependent media~\cite{Yariv-Yeh-book} was combined with the standard (but somewhat approximate) derivation of pulse propagation in optical fibers~\cite{Agrawal-book}. This approach, however, incorporated only material dispersion, and neglected structural dispersion.  Thus, similarly to~\cite{temporal_CMT_Hardy,Pile_on_Hardy}, this approach is limited to large structures, for which structural dispersion is negligible, so that pulse propagation in time-varying structures having (sub-)wavelength scale structure cannot be treated with the formalism of~\cite{temporal_CMT_Bahabad}. Moreover, the derivation of~\cite{temporal_CMT_Bahabad} is applied only to periodic perturbations (i.e., consisting of a discrete set of monochromatic waves) in the spirit of standard nonlinear wave mixing, i.e., it cannot be readily implemented for perturbations consisting of a localized temporal perturbation or a continuum of modes.

In this article, we extend the work of~\cite{temporal_CMT_Bahabad} and present what we believe is the first {\em exact} derivation of CMT for pulse propagation in dispersive media undergoing a spatio-temporal perturbation. Our approach enables treating material and especially structural dispersion to any degree of desired accuracy and allows for any 3D spatio-temporal perturbation profiles, including non-periodic and localized perturbations in both space and time. % {\bf It also improves upon the standard models for nonlinear pulse propagation. ??}

Remarkably, our approach is derived from first principles, i.e., it does not require any phenomenological additions, is much simpler than some previous derivations, e.g., those relying on reciprocity~\cite{Chuang_CMT,temporal_CMT_Panoiu_FWMix} and can be implemented with minimal additional complexity compared with some other previously published derivations~\cite{temporal_CMT_Bahabad}. Yet, it does not rely on \underline{any} approximation, thus, offering a simple and light alternative to FDTD which is exact, easy to code and quick to run. Indeed, we expect that as for the CW case, where the vast majority of studies refrain from solving the Helmholtz equations and rely on CMT instead, our approach would become commonly used in problems of pulse propagation in time-varying media. Since we deal below primarily with scalar wave equations, our results could also be used in many additional contexts such as acoustic waves~\cite{Fink_review_3,Armenians_sound_switching}, spin waves~\cite{magnonics-book}, water waves~\cite{Fink-Cauchy-problem}, quantum/matter waves~\cite{BEC-book,Mendonca-Guerreiro} etc..

Our formulation is general, however, it takes the spirit of a waveguide geometry - it assumes a preferred direction of propagation, which could later facilitate the use of the paraxial approximation. % {\bf KEEP? Indeed, the examples in Section~\ref{sec:valid} are all on the waveguide configuration. }
Nevertheless, the approach can be extended also to $z$ variant structures, e.g., photonic crystals~\cite{deep_gratings_PRE_96,deep_grating_2D,Winn_Fan_Joann_Ippen_interband,Sivan-Pendry-article,temporal_CMT_gratings_russians}, and can be applied in conjunction with more compact formalisms such as the unidirectional flux models~\cite{Uni-directionality,Biancalana_1D_layered_media_n(t)}.

%will help Zev? probably CW..

Our paper is organized as follows. In Section~\ref{sub:definition}, we formulate the problem and define the notations, and in Section~\ref{sub:derivation}, we present our derivation of the Coupled Mode Theory equations in detail. In Section~\ref{sub:leading}, we discuss the hierarchy of the various terms in our model equations, isolate the leading order terms and provide a simple analytical solution to the leading order equation, which involves only a trivial integral. In Section~\ref{sec:valid}, we validate our derivation and its leading order solution using numerical simulations based on the FDTD method. % To the best of our knowledge, this is the first validation of the CMT for pulse propagation and for spatio-temporal perturbations.
Specifically, we study the interaction of several pulses of different temporal and spatial extents. We demonstrate the flexible control of the generated pulse duration, including spatio-temporal broadening or compression, via a complex, broadband wave mixing process which involves the exchange of spectral components between the interacting pulses. We believe that such results are described here for the first time. Finally, in Section~\ref{sec:discussion}, we discuss the importance, advantages and weaknesses of our results and compare them to previous derivations. For ease of reading, many of the details of the calculations are deferred to the Appendices.

%To Do:
%\begin{itemize}
%%   \item self action - can we explain coupling to higher order modes in fibers? FIX ERROR!!
%%   \item App. B + example - maybe thick optical fiber with low contrast + coupling to higher modes - maybe can ask Shai to calculate this with the weakly guiding example?
%\end{itemize}

\section{Derivation of the coupled mode equations for pulses in time varying media}\label{sec:derivation}
\subsection{Definition of the problem}\label{sub:definition}
{\color{black} For consistency and clarity of notation, all properties that have an explicit frequency dependence appear in lower case, while properties in the time domain appear in upper case. If two functions are related via a Fourier transform, denoted by the Fourier transform symbol, $\mathcal{F}^\omega_t \equiv \int_{-\infty}^\infty dt e^{i \omega t}$ or the inverse transform $\mathcal{F}_\omega^t \equiv \frac{1}{2\pi}\int_{-\infty}^\infty d\omega e^{- i \omega t}$, then they simply share the same letter, e.g., $A_m(t) \equiv \mathcal{F}^t_\omega[a_m(\omega)]$\footnote{An exception will be made for the permittivity, $R(t) = \mathcal{F}^t_\omega[\epsilon]$, and the coupling coefficient $\mathcal{K}(t) = \mathcal{F}^t_\omega[\kappa]$, see below. }.

Here, we consider pulse propagation in media whose optical properties are invariant in the propagation direction, chosen to be the $z$-direction, but can have any sort of non-uniformity in the transverse direction $\vec{r}_\perp$, i.e., the dispersive permittivity of the structure (defined in the {\em frequency} domain) is given by $\epsilon = \epsilon(\vec{r}_\perp,\omega)$. This is the natural waveguide geometry, which obviously includes also uniform media. However, the formulation presented below can be extended to more complicated structures such as gratings~\cite{temporal_CMT_gratings_russians} or photonic crystals~\cite{deep_gratings_PRE_96,Euser_Thesis,Winn_Fan_Joann_Ippen_interband,Sivan-Pendry-letter,Sivan-Pendry-article} by using the proper Floquet-Bloch modes, or even to photonic crystal waveguides by averaging along the propagation direction over a unit cell~\cite{temporal_CMT_Panoiu_FWMix}.

For such configurations, it is useful to consider modes - field profiles that are invariant along the transverse coordinates and accumulate phase as they propagate in the $z$ direction. For simplicity, we consider TE wave propagation in a uniform slab waveguide, since in this case, $\nabla \cdot \vec{E} \equiv 0$, thus, simplifying the derivation by allowing to neglect vectorial coupling. The equations for TM or mixed polarization or for 3D geometries follow a similar procedure, see Appendix~\ref{app:vector_Helmholtz}, with the only difference being somewhat different coefficients for the various terms. This choice makes our derivation below more general - it could then be used for more general wave systems which can be described by scalar wave equations, such as acoustic waves~\cite{Fink_review_3}, spin waves~\cite{magnonics-book}, water waves~\cite{Fink-Cauchy-problem}, matter waves~\cite{BEC-book} etc..

In this case, the modes of the Helmholtz equation are given by $\vec{e}_m(\vec{r}_\perp,\omega) e^{- i \omega t + i \beta_m(\omega) z}$ where  $\vec{e}_m(\vec{r}_\perp,\omega)$ are solutions of
\begin{equation}\label{eq:E_m}
\left[\nabla^2_\perp + \omega^2 \mu \epsilon(\vec{r}_\perp,\omega) - \beta_m^2(\omega) \right]\vec{e}_m(\vec{r}_\perp,\omega) = 0,
\end{equation}
where $\beta_m$ is the propagation constant of mode $m$, accounting for both structural and material dispersion. The mode normalization is taken in the standard way~\cite{Yariv-Yeh-book}, i.e.,
\begin{equation}\label{eq:normalization}
\int \int \vec{e}_m^*(\vec{r}_\perp,\omega) \cdot \vec{e}_n(\vec{r}_\perp,\omega) dx dy = \frac{2 \mu_0 \omega}{|\beta_m(\omega)|} \delta_{m,n},
\end{equation}
where $\mu_0 $ is the vacuum permeability and $\delta_{m,n}$ is a Kronecker's delta. Note that Eqs.~(\ref{eq:E_m})-(\ref{eq:normalization}) apply for both bound and radiative modes~\cite{Menyuk-radiative-modes}.

Under these conditions, Maxwell's equations reduce to
\begin{equation}\label{eq:vector_waeq}
\nabla^2 \vec{E}(\vec{r}_\perp,z,t) = \mu \frac{\partial^2}{\partial t^2} \vec{D}(\vec{r}_\perp,z,t),
\end{equation}
where
\begin{equation}\label{eq:D}
\vec{D}(\vec{r}_\perp,z,t) = \epsilon_0 \vec{E}(\vec{r}_\perp,z,t) + \vec{P}(\vec{r}_\perp,z,t),
\end{equation}
and where
\begin{equation}\label{eq:P}
\vec{P}(\vec{r}_\perp,z,t) = R(\vec{r}_\perp,t) \times \vec{E}(\vec{r}_\perp,z,t) = \mathcal{F}_\omega^t[\left(\epsilon(\vec{r}_\perp,\omega) - 1\right)\vec{e}(\vec{r}_\perp,z,\omega)].
\end{equation}
Here, $R(\vec{r}_\perp,t) \equiv \mathcal{F}_\omega^t[\epsilon(\vec{r}_\perp,\omega) - 1]$ is the response function of the media, representing its optical memory, and $\vec{e}(\vec{r}_\perp,z,\omega) \equiv \mathcal{F}^\omega_t(\vec{E}(\vec{r}_\perp,z,t))$. Note that $\times$ stands for a convolution.

In order to account for dispersion in a time-domain formulation (e.g., Finite Differences Time Domain, (FDTD)), the material polarization $\vec{P}$ is calculated self-consistently by coupling the Maxwell equations to an auxiliary differential equation, as shown in~\cite{Taflove-book} for linear media with both free and bound charge carriers. In media with a cubic nonlinearity, e.g., Kerr % (which is essentially instantaneous (or in other words, dispersionless))
or Raman nonlinearities~\cite{Taflove-book} or even free carrier nonlinearities~\cite{FDTD_FC_Agrawal_2010}, this procedure was applied as well, while still taking into account dispersion without any approximation. Similarly, one can account for other nonlinearities, such as thermal stress~\cite{Reed_doppler,Reed_shock} or thermal nonlinearities~\cite{Biancalana_NJP_2012,Stoll_review}, or more generally, for phenomenological changes of the permittivity. In these cases, the nonlinearity is again calculated exactly via an auxiliary differential equation, but eventually appear in the Maxwell equations as a {\em multiplicative} factor, i.e.,
\begin{equation}\label{eq:P_NL}
\Delta\vec{P}(\vec{r}_\perp,z,t) = \Delta R(\vec{r}_\perp,z,t) \vec{E}(\vec{r}_\perp,z,t),
\end{equation}
Here, $\Delta R(\vec{r}_\perp,z,t)$ is the spatio-{\em temporal} perturbation of the permittivity which is the Fourier transform of the perturbation of the permittivity, i.e.,
\begin{equation}\label{eq:delta_eps_hat}
\Delta\epsilon(\vec{r}_\perp,z,\omega) = \mathcal{F}_t^\omega [\Delta R(\vec{r}_\perp,z,t)].
\end{equation}
Note that in Eqs.~(\ref{eq:P})-(\ref{eq:P_NL}) we insist on denoting the response function and perturbation in the time domain by $R$ and $\Delta R$ (rather than $\epsilon(t)$ and $\Delta \epsilon(t)$, as done in many previous studies) in order to emphasize the difference between the frequency dependence of the material response, which, by definition is called {\em dispersion}, and the direct time dependence of this response due to temporal variation of the material properties themselves (such as free-carrier density, temperature etc.). Somewhat confusingly, frequently the latter effect is also referred to as dispersion. This notational distinction also serves to emphasize the fact that dispersion of the perturbation is fully taken into account, unlike several previous studies. We also assume here that $\Delta R$ is a scalar function; the extension to the vectorial case is tedious, and does not change any of the main features of the current derivation.

In Eq.~(\ref{eq:P_NL}) we treat the nonlinear terms as a spatio-temporal perturbation. Indeed, one can also introduce a perturbation in a phenomenological way using the same relation. It may distort the incoming mode but also couple it to additional modes. Note that unlike the unperturbed permittivity $\epsilon(\vec{r}_\perp,\omega)$, which is $z$-invariant, in the current derivation, the perturbation can have {\em any} spatio-temporal profile, i.e., $\Delta \epsilon(\vec{r}_\perp,z,\omega)$. Thus, in what follows we solve
\begin{equation}\label{eq:vector_waeq_pert}
\nabla^2 \vec{E}(\vec{r}_\perp,z,t) = \mu \frac{\partial^2}{\partial t^2} \left(\vec{P}(\vec{r}_\perp,z,t) + \Delta \vec{P}(\vec{r}_\perp,z,t)\right),
\end{equation}
where $\vec{P}$ and $\Delta \vec{P}$ are given by Eqs.~(\ref{eq:P})-(\ref{eq:P_NL}), respectively. In order to keep our results general, in what follows, we assume a general functional form for $\Delta R$ without dwelling into its exact source.

Finally, the formulation below can be applied also to $\chi^{(2)}$ media, if only the customary frequency domain definitions of the second order polarization are properly adapted to the time domain and the nonlinear polarization is written in the form of the perturbation used in~(\ref{eq:P_NL}).

\subsection{Derivation}\label{sub:derivation}
We assume that the total field is pulsed and decompose it as a discrete set of continuous sums of modes, {\em each with its own slowly-varying modal amplitude $a_m$ and wavevector $\beta_m(\omega)$}, namely,~\footnote{Note that in what follows, we implicitly assume everywhere that the real part of all relevant expressions is taken when interpreting physical quantities. }
\begin{subequations}
\begin{eqnarray}
\vec{E}(\vec{r}_\perp,z,t) &=& Re\left(\sum_m \mathcal{F}^t_\omega \left[\tilde{a}_m(z,\omega - \omega_0) e^{i \beta_m(\omega) z} \vec{e}_m(\vec{r}_\perp,\omega)\right]\right) \label{eq:E_ansatz1} \\
&=& Re\left(\sum_m e^{i \beta_{m,0} z} \mathcal{F}^t_\omega \left[a_m(z,\omega - \omega_0) \vec{e}_m(\vec{r}_\perp,\omega)\right]\right), \label{eq:E_ansatz2}
\end{eqnarray}
\end{subequations}
where $\beta_{m,0} \equiv \beta_m(\omega_0)$ and
\begin{equation}\label{eq:A_m_hat_tilde}
a(z,\omega - \omega_0) = \tilde{a}(z,\omega - \omega_0) e^{i (\beta_m(\omega) - \beta_{m,0})z},
\end{equation}
i.e., in Eq.~(\ref{eq:E_ansatz2}) we separated the (rapidly-varying) frequency-{\em in}dependent part of the dispersion exponent from the (slowly-varying) frequency-dependent part. Below we refer to each of the $m$ components of the sum in Eqs.~(\ref{eq:E_ansatz1})-(\ref{eq:E_ansatz2}) as a pulsed wave packet.

We note that the ansatz~(\ref{eq:E_ansatz1}) is the natural extension of the ansatz used to describe pulse propagation in {\em homogeneous, linear} media, see e.g.,~\cite{Jackson-book}, but differs from the approach taken in standard perturbation theory~\cite{temporal_CMT_Bahabad,temporal_CMT_Panoiu}, see Appendix~\ref{app:standard_derivations}, or in bulk Kerr media~\cite{Agrawal-book,Fibich-book}. Also note that the ansatz~(\ref{eq:E_ansatz1}) is the natural extension of the ansatz used in standard CMT formalism, which is applied to monochromatic waves (i.e., for $a_m(z,\omega - \omega_0) = A_m(z) \delta(\omega - \omega_0) + A_m^*(z) \delta(\omega + \omega_0)$, where $\omega_0$ is the carrier wave frequency), see e.g.,~\cite{Yariv-Yeh-book}. In this case, the total electric field in the time domain is simply given by
\begin{equation}\label{eq:E_CW}
\vec{E}(\vec{r}_\perp,z,t) = e^{- i \omega_0 t} \sum_m e^{i \beta_{m,0} z} A_m(z) \vec{e}_m(\vec{r}_\perp,\omega_0),
\end{equation}
so that $A_m(z)$ is the modal amplitude (the difference between $\tilde{a}_m(\omega)$ and $a_m(\omega)$ does not come into play in this case). In contrast, in the current context, for which the spectral amplitude $\tilde{a}_m$ has a {\em finite} width, $\vec{E}$~(\ref{eq:E_ansatz2}) is given % by a convolution of the mode profile $\vec{e}_m$, the slowly-varying modal amplitude $\tilde{a}_m$ and the exponent containing the full information about the dispersion in the system $e^{i \beta_m(\omega) z}$. In this case, one can
as the Fourier transform of the product of $a_m(z,\omega - \omega_0)$ and $\vec{e}_m(\vec{r}_\perp,\omega)$, so that it can also be written as
\begin{eqnarray}\label{eq:E_pulse}
\vec{E}(\vec{r}_\perp,z,t) &=& \sum_m e^{i \beta_{m,0} z} \mathcal{F}^t_\omega\left[ \int dt' A_m(z,t') e^{i \omega t'} \int dt'' \mathcal{F}_\omega^{t''}[\vec{e}_m(\vec{r}_\perp,\omega)] e^{i \omega t''}\right] \nn \\
&=& \sum_m e^{i \beta_{m,0} z} \int d\tau A_m(z,t - \tau) e^{i \omega_0 \tau} \mathcal{F}_\omega^\tau[\vec{e}_m(\vec{r}_\perp,\omega)] \nn \\
&=& \sum_m e^{- i \omega_0 t + i \beta_{m,0} z} \int d\tau A_m(z,t - \tau) e^{i \omega_0 (\tau + t)} \mathcal{F}_\omega^\tau[\vec{e}_m(\vec{r}_\perp,\omega)], \end{eqnarray}
where, following previous definitions, we used
\begin{eqnarray}\label{eq:A_m}
A_m(z,t) &\equiv& \mathcal{F}_\omega^t \left[a_m(z,\omega - \omega_0)\right] \equiv \frac{1}{2\pi} \int d\omega e^{- i \omega t} a_m(z,\omega - \omega_0) \nn \\
&=& \frac{1}{2\pi} e^{- i \omega_0 t} \int d\omega e^{- i (\omega - \omega_0)t } a_m(z,\omega - \omega_0).
\end{eqnarray}
Thus, strictly speaking, the total electric field {\em cannot} be written as a simple product of a mode with carrier frequency $\omega_0$ and a slowly-varying amplitude, see Appendix~\ref{app:standard_derivations}, unless dispersion is unimportant~\cite{temporal_CMT_Hardy}. Yet, since the mode profile $\vec{e}_m(\vec{r}_\perp,\omega)$ is generically very wide spectrally, the spectral mode amplitude $\tilde{a}_m$ is effectively a delta function, so that to a very good approximation, the convolution~(\ref{eq:E_pulse}) reduces to the product $A_m(z,t)\ \vec{e}_m(\vec{r}_\perp,\omega_0)$. We believe that already this subtle point was not noted in previous derivations; it will become important for pulses of decreasing duration.

When the system includes pulses centered at well-separated frequencies (as e.g., in~\cite{temporal_CMT_Bahabad}), then, one needs to replace $\omega_0$ by $\omega_p$ and sum over the modes $p$, see below~\footnote{In this context, Eq.~(\ref{eq:E_ansatz1}) should formally include also the complex conjugate of the term on the LHS. This term would come into play only for Four-wave mixing interactions in Kerr media.
}.

Now, using the ansatz~(\ref{eq:E_ansatz1}) and Eqs.~(\ref{eq:P})-(\ref{eq:delta_eps_hat}) in Eq.~(\ref{eq:vector_waeq_pert}) leads to the cancellation of the terms accounting for transverse diffraction, material and structural dispersion for {\em all} frequencies, see comparison to standard derivations in Appendix~\ref{app:standard_derivations}. We are thus left with
\begin{eqnarray}\label{eq:CMT1}
&& \int_{-\infty}^\infty d\omega e^{- i \omega t} \sum_m e^{i \beta_m(\omega) z} \left[\frac{\partial^2}{\partial z^2} \tilde{a}_m(z,\omega - \omega_0) + 2 i \beta_m(\omega) \frac{\partial}{\partial z} \tilde{a}_m(z,\omega - \omega_0)\right] \vec{e}_m(\vec{r}_\perp,\omega) \nn  \\
&& \quad = \mu \frac{\partial^2}{\partial t^2} \left[\left(\int_{-\infty}^\infty d\omega e^{- i \omega t} \Delta\epsilon(\vec{r}_\perp,z,\omega)\right) \frac{1}{2\pi} \left(\sum_n \int_{-\infty}^\infty d\omega e^{- i \omega t} e^{i \beta_n(\omega) z} \tilde{a}_n(z,\omega - \omega_0) \vec{e}_n(\vec{r}_\perp,\omega)\right)\right] \nn \\
&& \quad = - \mu \int_{-\infty}^\infty d\omega \omega^2 e^{- i \omega t} \Delta \vec{p}(\omega),
\end{eqnarray}
where
\begin{equation}\label{eq:dp}
\Delta \vec{p}(\omega) = \sum_n \int_{-\infty}^\infty d\omega' \Delta\epsilon(\vec{r}_\perp,z,\omega') e^{i \beta_n(\omega - \omega') z} \tilde{a}_n(z,\omega - \omega' - \omega_0) \vec{e}_n(\vec{r}_\perp,\omega - \omega').
\end{equation}

For simplicity, we suppress the Fourier transform symbol ($\int_{-\infty}^\infty d\omega e^{- i \omega t}$) from both sides of Eq.~(\ref{eq:CMT1}). We proceed by taking the scalar product of Eq.~(\ref{eq:dp}) with $\vec{e}^*_m(\vec{r}_\perp,\omega)$, integrating over $\vec{r}_\perp$ and dividing by the modal norm~(\ref{eq:normalization}). This yields
\begin{eqnarray}\label{eq:CMT2}
\!\!\!\! e^{i (\beta_m(\omega) - \beta_{m,0})z} \left[\frac{\partial^2}{\partial z^2} \tilde{a}_m(z,\omega - \omega_0) + 2 i \beta_m(\omega) \frac{\partial}{\partial z} \tilde{a}_m\right] = - \frac{\omega |\beta_m(\omega)|}{2} e^{- i \beta_{m,0} z} \sum_n c_{m,n}(z,\omega,\omega - \omega_0), \nn \\
\end{eqnarray}
where we defined
\begin{eqnarray}\label{eq:c}
c_{m,n}(z,\omega,\omega - \omega_0) = \int d\omega' \kappa_{m,n}(z,\omega,\omega') e^{i \beta_n(\omega - \omega') z} \tilde{a}_n(z,\omega - \omega' - \omega_0),
\end{eqnarray}
and the mode coupling coefficient
\begin{equation}\label{eq:kappa}
\kappa_{m,n}(z,\omega,\omega') = \int d\vec{r}_\perp \left[\vec{e}_m^*(\vec{r}_\perp,\omega) \cdot \vec{e}_n(\vec{r}_\perp,\omega - \omega')\right] \Delta \epsilon(\vec{r}_\perp,z,\omega').
\end{equation}
As noted, for a mode profile that satisfies the vector Helmholtz equation (rather than the (regular) Helmholtz equation that is assumed here), only small changes to the weights of the various terms will be incurred, see Appendix~\ref{app:vector_Helmholtz}.

Note that Eq.~(\ref{eq:CMT2}) is an integro-differential equation, which can be simplified in several limits. However, we prefer to follow a more general approach in which we derive a purely differential model in the time domain. In order to do that, we Fourier transform Eq.~(\ref{eq:CMT2}) (see Appendix~\ref{app:FT} for details). This yields
\begin{eqnarray}\label{eq:CMT_t}
&& \frac{\partial^2}{\partial z^2} A_m(z,t) + 2 i \beta_{m,0} \frac{\partial}{\partial z} A_m + 2 i \frac{\beta_{m,0}}{v_{g,m}} \frac{\partial}{\partial t} A_m - \left(\frac{1}{v_{g,m}^2} + \beta_{m,0} \beta''_{m,0}\right) \frac{\partial^2}{\partial t^2} A_m + \sum_{q=3}^\infty \alpha_{q,m} \left(i\frac{\partial}{\partial t}\right)^q A_m \nn \\
&& \quad \quad = - \frac{\omega_0 |\beta_{m,0}|}{2} \sum_n e^{i (\beta_{n,0} - \beta_{m,0}) z}\mathcal{K}_{m,n}(z,\omega_0,t) A_n(z,t) + H.O.T,
\end{eqnarray}
where
\begin{equation}\label{eq:v_gm}
v_{g,m} = \left(\frac{\partial \beta_m(\omega)}{\partial \omega}\bigg|_{\omega = \omega_0}\right)^{-1},
\end{equation}
is the group velocity of mode $m$, $\beta''_{m,0} \equiv \partial^2 \beta_m(\omega) / \partial \omega^2|_{\omega = \omega_0}$, $\alpha_{q,m} \equiv \frac{1}{q!}\frac{d^q (\beta_m^2)}{d \omega^q}\big|_{\omega_0}$ are the higher-order dispersion coefficients, and H.O.T stands for the high-order dispersive terms of the perturbation. Note that the group velocity dispersion (GVD) term (proportional to $\partial^2 / \partial t^2$), as well as the higher-order dispersion terms, on the LHS of Eq.~(\ref{eq:CMT_t}) include a term that will be eliminated once a transformation to a coordinate frame co-moving with the pulse is performed~\cite{Jain-Tzoar,Fibich-book} at the cost of an additional term $A_{zt}$, responsible for spatio-temporal coupling. This term is missing in most standard derivations~\cite{Agrawal-book}.

The coupling coefficient $\mathcal{K}$ is defined as (see Appendix~\ref{app:FT}})
\begin{eqnarray}\label{eq:kappa_t}
\mathcal{K}_{m,n}(z,\omega_0,t) &=& \mathcal{F}_{\omega'}^t[\kappa_{m,n}(z,\omega_0,\omega')] % = \int d\omega' \kappa_{m,n}(z,\omega_0,\omega') e^{- i \omega' t} \nn
\\ &=& \int d\omega' \int d\vec{r}_\perp \Delta \epsilon(\vec{r}_\perp,z,\omega') \left[\vec{e}_m^*(\vec{r}_\perp,\omega_0) \cdot \vec{e}_n(\vec{r}_\perp,\omega_0 - \omega')\right] e^{- i \omega' t}. \nn
\end{eqnarray}
Eq.~(\ref{eq:kappa_t}) shows that the coupling coefficient involves spatial averaging of mixed spectral components. More generally, the coupling coefficient shows that the spatio-temporal perturbation manifests an unusual coupling between the (transverse) spatial (via the $z$ dependence) and the temporal (via the $t$ dependence) degrees of freedom of the perturbing pump pulse.

For simplicity, let us assume that the dependence of the perturbation on the transverse coordinates $\vec{r}_\perp$ is separable from the dependence on the other coordinates, or in other words, that the transverse variation of the perturbation is weakly dependent on the frequency or location along the waveguide, i.e., $\Delta \epsilon(\vec{r}_\perp,z,\omega') = \Delta \epsilon(z,\omega') W(\vec{r}_\perp)$; this is a good assumption for a thin waveguide. In this case,
\begin{eqnarray}\label{eq:kappa_t_approx}
\mathcal{K}_{m,n}(z,\omega_0,t) = \int d\omega' \Delta \epsilon(z,\omega') o_{m,n}(\omega_0,\omega') e^{- i \omega' t} = \Delta R(z,t) \times O_{m,n}(\omega_0,t),
\end{eqnarray}
where we defined the overlap function
\begin{eqnarray}\label{eq:omn}
o_{m,n}(\omega_0,\omega) = \int d\vec{r}_\perp W(\vec{r}_\perp) \vec{e}_m^*(\vec{r}_\perp,\omega_0) \cdot \vec{e}_n(\vec{r}_\perp,\omega_0 - \omega).
\end{eqnarray}
The overlap integral~(\ref{eq:omn}) is an extension of the well-known coupling coefficient appearing in the standard CMT for monochromatic waves~\cite{Yariv-Yeh-book}, as well as the one arising in the approximate CMT for pulses under a periodic perturbation, see e.g.,~\cite{temporal_CMT_Bahabad}. This is another difference of our exact derivation with respect to previous derivations. However, the overlap function is generically, slowly-varying with the frequency, % {\bf I expect the f to f (and b??) conversion not to be so wide spectrally because once the momenta shift, I expect the overlap to drop quickly..!}
so that we can Taylor expand it around $\omega' = 0$, and get
\begin{eqnarray}\label{eq:kappa_t_approx2}
\mathcal{K}_{m,n}(z,\omega_0,t) &=& o_{m,n}(\omega_0,0) \Delta \epsilon(z,t) + i \frac{\partial o_{m,n}(\omega_0,\omega)}{\partial \omega}\bigg|_{\omega = 0} \frac{\partial}{\partial t} \Delta \epsilon(z,t) + \cdots.
\end{eqnarray}
Thus, by neglecting the dispersion of the perturbation, the familiar expression is retrieved, together with the ``adiabatic'' time variation of the refractive index. This result justifies, a-posteriory, the many instances where such expressions were used without a formal proof.

Eq.~(\ref{eq:CMT_t}) is our main result - it is an equation for the ``mode amplitude'' in the time domain ($A_m(t)$, see Eqs.~(\ref{eq:E_pulse})-(\ref{eq:A_m})) where the dependence on the fast oscillations associated with the optical cycle was removed. This formulation facilitates easy coding and quick and light numerical simulations. Note that if the total field includes additional pulses centered at different frequencies, then, an additional phase-mismatch term associated with frequency detuning will be added to the exponent in Eq.~(\ref{eq:CMT_t}), see~\cite{Bahabad_NP,temporal_CMT_Bahabad}.

\subsection{The leading order equation}\label{sub:leading}

We now would like to classify the relative magnitude of the various terms in Eq.~(\ref{eq:CMT_t}). In the standard case of pulse propagation in media with time-{\em in}dependent optical properties, there are two time scales - the period $T_0 = 2\pi/\omega_0$ and $T_m$, the duration of the pulse $m$~\cite{Agrawal-book,Fibich-book}. However, in the presence of a perturbation which is localized in space and time, there are two additional time scales, associated with the duration and the spatial extent of the perturbation, namely, the switching time $T_{sw}$ and the time $T_{pass} = L/v_g$ it takes a photon to cross the spatial extent of the perturbation $L$, respectively (see also Section~\ref{sec:valid} below).

Due to this complexity, it is not possible to perform a general classification of the relative magnitude of the various terms in Eq.~(\ref{eq:CMT_t}). However, in many cases, the time (length) scales associated with the pulse duration and perturbation are at least several cycles (wavelengths) long, so that the non-paraxiality term is negligible compared with both the second and third terms on the LHS of Eq.~(\ref{eq:CMT_t}), i.e., $\frac{\partial^2}{\partial z^2} A_m(z,t) \sim (v_g T_m)^{-2} \ll \beta_{m,0} \frac{\partial}{\partial z} A_m \sim \beta_{m,0} v_g^{-1} \frac{\partial}{\partial t} A_m \sim \beta_m (v_g T_m)^{-1}$ .

In addition, in most cases, only $A_{f}(z,t)$, corresponding to the forward (i.e., $v_{g,f} > 0$) propagating fundamental bound mode of the waveguide at $\omega_0$ and wavevector $\beta_{m,0} = \beta_f > 0$, does not vanish for $t \to - \infty$. In this case, a Maclaurin expansion of the amplitudes $A_m(z,t)$ in $\Delta \epsilon$, it is clear that $A_{f}(z,t)$ is the only amplitude that does not vanish in zero-order in $\Delta \epsilon$ for all $t$, and all other modes are at least $O(\Delta \epsilon)$ small. In these cases, the coefficient $\kappa_{m,\pm m}$ can be approximated using Eq.~(\ref{eq:normalization}) as $\omega \Delta \epsilon/c^2 \beta_m$, so that $\omega \beta_m \kappa_{m,m} A_f \sim \beta_m^2 \Delta \epsilon$. The contribution of the terms involving coupling between different modes is typically smaller by an extent depending on the details of the problem. In addition, unless structures with very strong dispersion are involved, such as e.g., photonic crystal waveguides~\cite{phc_wgs_slow_light,de_Rossi_pulse_compression_PhC_wgs} or plasmonic waveguides~\cite{plasmonic_grating_slow_vg,MIM_EIT,plasmonic_slow_vg,plasmonic_slow_vg_Kuipers}%\footnote{\bf near resonance? fast/slow light points!}
, all the additional terms, associated with time-derivatives of the mode amplitudes are typically smaller. Thus, in most cases, the group velocity dispersion (GVD) and high-order dispersion (HOD) terms can be neglected.

Combining all the above, and under the reasonable assumption that $\beta_m \Delta \epsilon \sim (v_g T_m)^{-1}$, we can keep terms only up to first-order on the LHS and zero-order on the RHS and obtain\footnote{Note the slight difference in notation compared to~\cite{Sivan-COPS-switching-TBG}. }
\begin{equation}\label{eq:CMT_approx_1}
\left[\frac{\partial}{\partial z} + \frac{1}{v_{g,m}} \frac{\partial}{\partial t}\right] A_m(z,t) = i \frac{\omega_0 sgn[\beta_{m,0}]}{4} \sum_n \mathcal{K}_{m,n}(z,\omega_0,t) e^{i \left(\beta_{n,0} - \beta_{m,0}\right)z} A_n(z,t).
\end{equation}
Keeping first order accuracy, the incoming mode amplitude is $A_{f}(z,t) = A_{f}^{inc}(z - v_{g,f} t)$ and the amplitudes of all other modes are given by
\begin{equation}\label{eq:CMT_approx_2}
\left[\frac{\partial}{\partial z} + \frac{1}{v_{g,m}} \frac{\partial}{\partial t}\right] A_m(z,t) = i \frac{\omega_0 sgn[\beta_{m,0}]}{4} \mathcal{K}_{m,f}(z,\omega_0,t) A_{f}^{inc}(z - v_{g,f} t) e^{i \left(\beta_f - \beta_{m,0}\right)z},
\end{equation}
where we set $\beta_f \equiv \beta_f(\omega_0)$. To proceed, we transform to a coordinate system co-propagating with the signal pulse, namely, transform from the variables $(z,t)$ to the variables ($z_m \equiv z - v_{g,m} t,t$). Then, after integrating over time~\footnote{In fact, as noted, one has to perform the transformation to the moving frame only before neglecting the non-paraxiality term, in which case, there is also the spatio-temporal coupling that has to be neglected in order to retrieve Eq.~(\ref{eq:CMT_approx_3}). }, we obtain
\begin{eqnarray}\label{eq:CMT_approx_3}
A_m(z_m,t) &=& i \frac{v_{g,m} \omega_0 sgn[\beta_{m,0}]}{4}  e^{i \left(\beta_{m,0} - \beta_f\right) z_m} \\
&& \int_{-\infty}^t \!\!\! d\tilde{t} \mathcal{K}_{m,f}(z_m - (v_{g,pert} - v_{g,m}) \tilde{t},\omega_0,\tilde{t}) A_{f}(z_m - (v_{g,f} - v_{g,m}) \tilde{t},\tilde{t}) e^{- i \left(\beta_{m,0} - \beta_f\right) v_{g,m} \tilde{t}}, \nn
\end{eqnarray}
where we assumed that the perturbation may move too (see e.g.,~\cite{Reed_doppler,Reed_shock,Winn_Fan_Joann_Ippen_interband,Fan_Lipson_interband_isolator,Philbin-Leonhardt-event-horizon,Notomi_review,Bahabad_NP}, an effect that is accounted for by writing the coupling coefficient as $\mathcal{K}_{m,n} = \mathcal{K}_{m,n}(z - v_{g,pert} t)$; similarly, one can account also for accelerating perturbations~\cite{Bahabad_accelerating}. Thus, after the perturbation is over, the amplitude of mode $m$ is given by a convolution of the perturbation and the incoming pulse, a result of the relative walk-off of the incoming forward pulse, the newly generated modes and the perturbation. The solution~(\ref{eq:CMT_approx_3}) generalizes the results of~\cite{Sivan-Pendry-letter,Sivan-Pendry-article,Sivan-Pendry-HSM}. % ,Sivan-short-pulses_TBG}.
We emphasize that this solution is rather simple - it applies for any perturbation, involves only a simple integral, and does not require any sophisticated asymptotics~\cite{Sivan-Pendry-article,Armenians_smooth_t_perturbation}.

\section{Validation}\label{sec:valid}

In order to validate the derivation of Eq.~(\ref{eq:CMT_t}), we compare the numerical solution of this model and its analytical solution~(\ref{eq:CMT_approx_3}) with exact FDTD simulations. The chosen example tests the new features of the expansion, namely, the quantitative comparison of the coupling under different temporal conditions. In that regard, we do not test the aspects related with the dispersion terms (on the LHS of Eq.~(\ref{eq:CMT_t})) as these are well understood.

% \newpage
% \subsection{Signal and pump pulses of comparable duration}
{\color{black} We consider an incoming forward propagating Gaussian pulse which has the transverse profile of the fundamental ({\em single}, for that instance) mode of the waveguide, namely,
\begin{equation}\label{eq:Gaussian-pulse}
E_f^{(inc)}(z,t) = \left[A_f^{(inc)}(z - v_{g,f} t) \vec{\bar{e}}_f(\vec{r}_\perp)\right] e^{- i \omega_0 t + i \beta_f z}, \quad \quad A_f^{(inc)} = e^{- \left(\frac{z - v_{g,f} t}{v_{g,f} T_f}\right)^2},
\end{equation}
where $T_f$ is the forward pulse initial duration, and $\vec{\bar{e}}_f \ne \vec{e}(\vec{r}_\perp,z,\omega) \ne \vec{e}_m(\vec{r}_\perp,\omega_0)$ is the transverse profile {\em of the wave packet}, both derived from Eq.~(\ref{eq:E_ansatz1}) and Eq.~(\ref{eq:E_pulse})~\footnote{Note that by Eq.~(\ref{eq:E_ansatz1}), this may not be always possible. }.

We now consider a case where the perturbation is a transient Bragg grating (TBG, see e.g.,~\cite{Sivan-COPS-switching-TBG}) of a finite length within the waveguide, namely,
\begin{equation}\label{eq:Delta_R_example1}
\Delta R(\vec{r}_\perp,z,t) = \epsilon_0 \Delta \bar{\epsilon} W(\vec{r}_\perp) q(z) e^{- \left(\frac{z}{L}\right)^2} e^{- \left(\frac{t}{T_{sw}}\right)^2},
\end{equation}
where
\begin{equation}\label{eq:q}
q(z) = \cos^2 (k_g z) = \frac{1 + \cos (2 k_g z)}{2},
\end{equation}
represents the periodic pattern of the TBG; note that it includes two Fourier components, $\pm 2 k_g$, but also a zero component;\footnote{Such a term is unavoidable with all nonlinear effects, but will be shown below to be negligible, at least to leading order.} $W$ is non-zero only within the waveguide and has unity magnitude, $L$ is the pump longitudinal length, and $T_{sw}$ is the characteristic time of the perturbation which can be shorter than the duration of an optical pump for a nonlinear interaction (e.g., a Kerr medium)%~\cite{Sivan-short-pulses_TBG}
, or longer for a temporally delayed nonlinearity like a free carrier nonlinearity~\cite{Sivan-COPS-switching-TBG} or a thermal nonlinearity. % ~\cite{Sivan-Spector}. % {\bf note there are in general 4 arbitrary time scales here - cycle, forward pulse duration, modulation duration and passage time! the latter three appear in the analysis. }
Note that $\Delta \bar{\epsilon} = max\left[\Delta R(\vec{r}_\perp,z,t)\right]$.
A Fourier transform of Eq.~(\ref{eq:Delta_R_example1}), we see that
\begin{equation}\label{eq:Delta_eps_example1}
\Delta \epsilon(\vec{r}_\perp,z,\omega) = \epsilon_0 \Delta \bar{\epsilon} \sqrt{\pi} T_{sw} W(\vec{r}_\perp) \frac{1 + \cos (2 k_g z)}{2} e^{- \left(\frac{z}{L}\right)^2} e^{- \left(\frac{T_{sw}}{2}\omega\right)^2}. \nn
\end{equation}
Now substituting in Eq.~(\ref{eq:kappa}) gives
\begin{eqnarray}\label{eq:kappa_example_1}
\kappa_{m,f}(z,\omega,\omega') = \epsilon_0 \sqrt{\pi} T_{sw} \Delta \bar{\epsilon} \frac{1 + \cos (2 k_g z)}{2} e^{- \left(\frac{z}{L}\right)^2} e^{- \left(\frac{T_{sw}}{2}\omega'\right)^2} \int_{-\infty}^\infty d\vec{r}_\perp W(\vec{r}_\perp) \left[\vec{e}_m^*(\vec{r}_\perp,\omega) \cdot \vec{e}_f(\vec{r}_\perp,\omega - \omega')\right]. \nn \\
\end{eqnarray}
For simplicity, we assume that the perturbation is sufficiently uniform, $W(\vec{r}_\perp) \approx 1$. This assumption is valid for a thin waveguide, and/or sufficiently long switching pulses and in the absence of substantial absorption in the waveguide material or reflections from its boundaries~\cite{Euser_Vos_JAP_2005}. In this case, an inverse Fourier transform of Eq.~(\ref{eq:kappa_example_1}) yields
\begin{eqnarray}
\mathcal{K}_{m,f}(z,\omega_0,t) &=& \frac{\epsilon_0 \sqrt{\pi}}{2\pi} T_{sw} \Delta \bar{\epsilon} \frac{1 + \cos (2 k_g z)}{2} e^{- \left(\frac{z}{L}\right)^2} \int_{-\infty}^\infty d\omega' e^{- \left(\frac{T_{sw}}{2}\omega'\right)^2} o_{m,f}(\omega_0,\omega') e^{- i \omega' t}, \nn \\ \label{eq:kappa_example_1a}
% o_{m,f}(\omega_0,\omega') &=& \int d\vec{r}_\perp \vec{e}_m^*(\vec{r}_\perp,\omega_0) \cdot \vec{e}_f(\vec{r}_\perp,\omega_0 - \omega'). \nn
\end{eqnarray}
where $o_{m,f}$ was defined in Eq.~(\ref{eq:omn}). Since the waveguide supports only a single mode, the only possible mode conversion is between the forward and backward waves\footnote{Note that we exclude here the possibility of coupling between modes of different polarizations or symmetries, since in the absence of an extremely non-uniform perturbation, the coupling between these modes is very weak. }. In this case, only the $- 2 k_g$ longitudinal Fourier component of the perturbation can lead to effective coupling, hence, we neglect the other (i.e., the $+ 2 k_g$ and $0$) Fourier components. In addition, we use approximation~(\ref{eq:kappa_t_approx2}) of the coupling coefficient, in which case, Eq.~(\ref{eq:kappa_example_1a}) becomes
\begin{eqnarray}\label{eq:kappa_example_1b}
\mathcal{K}_{b,f}(z,\omega_0,t) &\cong& \frac{\epsilon_0 \sqrt{\pi}}{2\pi} T_{sw} \Delta \bar{\epsilon} \frac{1 + \cos (2 k_g z)}{2} e^{- \left(\frac{z}{L}\right)^2} \int_{-\infty}^\infty d\omega' e^{- \left(\frac{T_{sw}}{2}\omega'\right)^2} o_{b,f}(\omega_0,0) e^{- i \omega' t} \nn \\
&\cong& \epsilon_0 \Delta \bar{\epsilon} o_{b,f}(\omega_0,0) \frac{e^{- 2 i k_g z}}{4} e^{- \left(\frac{z}{L}\right)^2} e^{- \left(\frac{t}{T_{sw}}\right)^2}. %  \nn \\
% &\cong& \frac{\Delta \bar{\epsilon} o_{b,f}}{4 c^2 \mu_0} \left[1 + \cos (2 k_g z)\right] e^{- \left(\frac{z}{L}\right)^2} e^{- \left(\frac{t}{T_{sw}}\right)^2}.
\end{eqnarray}
Substituting in Eq.~(\ref{eq:CMT_approx_3}) gives
\begin{eqnarray}\label{eq:A_b_1}
A_b(z_b,t) % &=& - i \frac{v_{g,b} \omega_0}{4} e^{i \left(\beta_b - \beta_f\right) z_b} \nn \\
% && \int_{-\infty}^t \!\!\! d\tilde{t} \mathcal{K}_{b,f}(z_b - (v_{g,pert} - v_{g,b}) \tilde{t},\omega_0,\tilde{t}) A_{f}(z_b - (v_{g,f} - v_{g,b}) \tilde{t},\tilde{t}) e^{- i \left(\beta_b - \beta_f\right) v_{g,b} \tilde{t}}, \nn \\
&=& - i \frac{v_g \omega_0}{4} e^{i \left(\beta_b - \beta_f\right) z_b} \frac{\epsilon_0 \Delta \bar{\epsilon} o_{b,f}(\omega_0,0)}{4} \nn \\
&& \int_{-\infty}^t \!\!\! d\tilde{t}  e^{- 2 i k_g (z_b - v_g \tilde{t})} e^{- \left(\frac{z_b - v_g \tilde{t}}{v_g T_{pass}}\right)^2} e^{- \left(\frac{\tilde{t}}{T_{sw}}\right)^2} e^{- \left(\frac{z_b - 2 v_g \tilde{t}}{v_g T_f}\right)^2} e^{- i \left(\beta_b - \beta_f\right) v_g \tilde{t}}, \nn
\end{eqnarray}
where $z_b = z + v_g t$ and we set $v_g = v_{g,f} = - v_{g,b}$ and as defined above, $L = v_g T_{pass}$. Choosing $\beta_f \equiv - \beta_b = k_g$, the expression for the backward pulse $A_b$ becomes % {\bf should be 16??}
\begin{eqnarray}\label{eq:A_b_2}
A_b(z_b,t) &=& - i \frac{\pi o_{b,f}(\omega_0,0)}{8 n_g \mu_0 \lambda_0} \Delta \bar{\epsilon} e^{2 i \beta_b z_b} \nn \\
&& \int_{-\infty}^t \!\!\! d\tilde{t} e^{- 2 i \beta_b (z_b - v_g \tilde{t})} e^{- \left(\frac{z_b - v_g \tilde{t}}{v_g T_{pass}}\right)^2} e^{- \left(\frac{\tilde{t}}{T_{sw}}\right)^2} e^{- \left(\frac{z_b - 2 v_g \tilde{t}}{v_g T_f}\right)^2} e^{- 2 i \beta_b v_g \tilde{t}} \nn \\
&\cong& - i \frac{\pi o_{b,f}(\omega_0,0)}{8 n_g \mu_0 \lambda_0} \Delta \bar{\epsilon} e^{2 i \beta_b z_b} \int_{-\infty}^t \!\!\! d\tilde{t} e^{- 2 i \beta_b (z_b - v_g \tilde{t})} e^{- \left(\frac{z_b - v_g \tilde{t}}{v_g T_{pass}}\right)^2} e^{- \left(\frac{\tilde{t}}{T_{sw}}\right)^2} e^{- \left(\frac{z_b - 2 v_g \tilde{t}}{v_g T_f}\right)^2} e^{- 2 i \beta_b v_g \tilde{t}} \nn \\
% &=& - i \frac{\omega_0}{8 n_g n_{f,eff}(\omega_0)} \Delta \bar{\epsilon} \int_{-\infty}^t \!\!\! d\tilde{t} e^{- \left(\frac{z_b - v_g \tilde{t}}{v_g T_{pass}}\right)^2} e^{- \left(\frac{\tilde{t}}{T_{sw}}\right)^2} e^{- \left(\frac{z_b - 2 v_g \tilde{t}}{v_g T_f}\right)^2} \nn \\
% &=& - i \frac{\omega_0}{8 n_g n_{f,eff}(\omega_0)} \Delta \bar{\epsilon} \int_{-\infty}^t \!\!\! d\tilde{t} e^{- \left(\frac{z_b^2 - 2 v_g \tilde{t} z_b + (v_g \tilde{t})^2}{(v_g T_{pass})^2}\right) - \left(\frac{\tilde{t}}{T_{sw}}\right)^2 - \left(\frac{z_b^2 - 4 v_g \tilde{t} z_b + (2 v_g \tilde{t})^2}{(v_g T_f)^2} \right)} \nn \\
&=& ... = - i \frac{\pi o_{b,f}(\omega_0,0)}{8 n_g \mu_0 \lambda_0} \Delta \bar{\epsilon} e^{- \frac{z_b^2}{(v_g T_1)^2}} \int_{-\infty}^t \!\!\! d\tilde{t} e^{- \frac{\tilde{t}^2}{T_4^2} + 2 \frac{z_b}{v_g T_2^2}\tilde{t}},
\end{eqnarray}
where $\lambda_0$ is the free space wavelength and
\begin{equation}\label{eq:Ts}
\frac{1}{T_4^2} = \frac{1}{T_{pass}^2} + \frac{1}{T_{sw}^2} + \frac{4}{T_f^2}, \quad \quad \frac{1}{T_2^2} = \frac{1}{T_{pass}^2} + \frac{2}{T_f^2}, \quad \quad \frac{1}{T_1^2} = \frac{1}{T_{pass}^2} + \frac{1}{T_f^2}. \nn
\end{equation}
Note that $T_1 > T_2 > T_4$. Taking the limit $t \to \infty$, it follows that
%\begin{eqnarray}% \label{eq:A_b}
%A_b(z,t) &=& - i \sqrt{\pi} \frac{\omega_0}{8 n_g n_{f,eff}(\omega_0)} \Delta \bar{\epsilon} T_4 e^{- \frac{z_b^2}{v_g^2}\left(\frac{1}{T_1^2} - \frac{T_4^2}{ T_2^4}\right)}. \nn
%\end{eqnarray}
%If we do not replace $o_{b,f}$ by its approximate value, the final results reads
\begin{eqnarray}\label{eq:A_b}
A_b(z,t) &=& - i \sqrt{\pi} \frac{\pi o_{b,f}(\omega_0,0)}{8 \mu_0 \lambda_0 n_g} \Delta \bar{\epsilon} T_4 e^{- \frac{\left(z + v_g t\right)^2}{v_g^2}\left(\frac{1}{T_1^2} - \frac{T_4^2}{ T_2^4}\right)}. %  \nn
\end{eqnarray}
% The solution~(\ref{eq:A_b}) is a generalization of results obtained in our previous studies. Indeed, in the limit of $T_{pass} \to \infty$ we get the results of~\cite{Sivan-Pendry-HSM} while in the limit $T_f \to \infty$, we retrieve the results of the short pulse generation ms~\cite{Sivan-short-pulses_TBG}.
{\color{black} The solution~(\ref{eq:A_b}) is a generalization of results obtained in~\cite{Sivan-Pendry-HSM} in the limit of $T_{pass} \to \infty$. It reveals an unusual coupling between the (transverse) spatial and temporal degrees of freedom of the perturbing pump pulse, represented by $T_{pass}$ and $T_{sw}$, respectively. It also shows that there is a natural trade-off between efficiency and duration - shorter interaction times/lengths give rise to shorter but weaker backward pulses. It can also be readily applied to other periodic perturbations, such as long Bragg gratings used for mode conversion~\cite{Fallnich-Hellwig}.

In order to validate the derivation of the CMT equation~(\ref{eq:CMT_t}), its leading order form~(\ref{eq:CMT_approx_2}) and its analytical solution~(\ref{eq:A_b}), we compare them with the results of an {\em exact FDTD} numerical solution of Maxwell equations using the commercial software package Lumerical Inc.~\cite{Lumerical}. Specifically, we consider a pulsed signal propagating in a single mode silica slab waveguide % with $n_{wg} = 1.5$ and $n_{clad} = 1.42$ having $2\mu$m thickness,
illuminated transversely by two interfering pulses of finite transverse extent, see Fig.~\ref{fig:configuration}. The interference period is chosen such that it introduces to the system the necessary momentum to couple efficiently the forward wave to the backward wave; no other (bound) modes are supported by the waveguide, hence, we account only for the coupling between these two (bound) modes. Such configuration is well known as a transient Bragg grating (TBG)~\cite{TBG-book}; its temporal and spatial extents and period can be easily controlled by adjusting the corresponding extents of the pump pulses and the angle between them.

In Fig.~\ref{fig:CMT_t_1}, we show the normalized power and duration of the backward pulse as a function of $T_{sw}$. In all the cases studied, all pulses maintain a Gaussian profile, and excellent agreement (agreement at 3 significant digits, at least) between the FDTD and CMT results is observed.
% {\bf discrepancies are potentially due to vectoral effect (not trivial that it would indeed be small..), GVD (not tested, not likely), efficiency (no), XPM (no), NP (not likely), coupling dispersion (not likely).. currently - analysis is 15\% higher.. }
Specifically, for pump intensities that correspond to maximal refractive index change of $\Delta n = 4 \cdot 10^{-3}$ via the Kerr nonlinearity of the silica, the backward pulse relative power is small ($< 1\%$), thus, the first order analytical solution~(\ref{eq:A_b}) is valid; it is indeed found to be in excellent agreement with the numerical results. { %\color{red} %{\bf NEED TO DISCUSS DIFFERENCES? MAYBE ONLY FOR HIGH EFFICIENCIES - gradual turn on, error in initial condition, signal depletion..}
Importantly, the duration of the generated backward pulse can be set to be either longer (temporal broadening) or even {\em shorter} (temporal compression) than the durations of all other time scales in the system. Indeed, this occurs because of a peculiar, {\em broadband} wave mixing process that involves the transfer of frequency components from the pump to the backward wave. To the best of our knowledge, such a spectral exchange process, in particular, of pulse compression, is presented here for the first time. In that regard, the generated backward pulse is not just a reflection - the current configuration reveals a rather simple, rich and somewhat non-intuitive way to control the intensity and duration of the generated (backward) pulse. Clearly, for signal and pump pulses of more complicated spatio-temporal profiles, this configuration opens the way to a novel and flexible way to shape the generated (backward and forward) pulse.

Finally, in order to demonstrate the power of our approach, we study the same configuration for a waveguide material that has a stronger optical nonlinearity, namely, the free-carrier (FC) nonlinearity; a prototypical example is a semiconductor, like silicon, SiN, GaAs etc.. On the conceptual level, such a nonlinearity cannot be treated with standard commercially available FDTD software, as one needs to couple the Maxwell equations to the rate equations governing the generation of free-carriers, involving single and multi-photon absorption, free-carrier diffusion and recombination, see e.g.,~\cite{Agrawal_Painter_review_OE_2007,FDTD_FC_Agrawal_2010,temporal_CMT_Panoiu_FWMix,Sivan-COPS-switching-TBG}. In that regard, the CMT approach provides a far simpler alternative to model the pulse dynamics, which does not require heavy computing. Moreover, such a nonlinearity enables much higher backward pulse generation efficiency, hence, giving the configuration at hand practical importance. More importantly, and in contrast to the standard scenarios, the FC nonlinearity can occur on very fast time scales. Indeed, usually, the FC nonlinearity is considered to be substantially slower compared with the instantaneous Kerr nonlinearity studied in the previous example. However, as shown in~\cite{Sivan-COPS-switching-TBG}, the TBG configuration solves this problem - carrier diffusion is fast enough on the scale of the interference period, such that the grating contrast washes out on sub picosecond time scales, in turn, enabling  sub-picosecond switching times. Indeed, once the grating contrast vanishes, the optical functionality of the perturbation, namely, the reflectivity, can disappear well before the recombination is complete, i.e., before the semiconductor system reached its equilibrium state. Thus, this configuration enables us to enjoy the best of the two worlds - strong switching combined with sub-picosecond features.

Fig.~\ref{fig:diffusive_switching} shows that the generated backward pulse can be several times shorter than the incoming forward pulse (up to 5-fold shorter in the examples shown; more substantial shortening is also possible at the cost of lower total coupling efficiency); the opposite is also possible (not shown). Moreover, the backward pulse total relative power can reach substantial values, up to about several percent. In addition, the forward pulse is partially absorbed by the generated free-carries, and becomes slightly temporally broader. For even stronger perturbations, we observe the formation of a more complicated backward pulse, with increasing number of side lobes in its trailing edge, while the forward wave develops a deep minimum where the backward pulse was extracted. Further investigation of this complex dynamics is deferred to a future investigation.

}
% a detailed discussion of the physics + intuition is provided in ...

}

\begin{figure}[htbp]
  \centering{\includegraphics[scale=0.8]{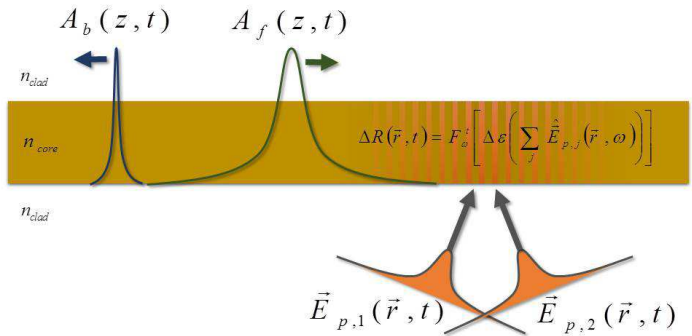}}
  \caption[]{(Color online) Schematic illustration of the configuration of the transient Bragg grating (TBG). } \label{fig:configuration}
\end{figure}

\begin{figure}[htbp]
  \centering{\includegraphics[scale=1]{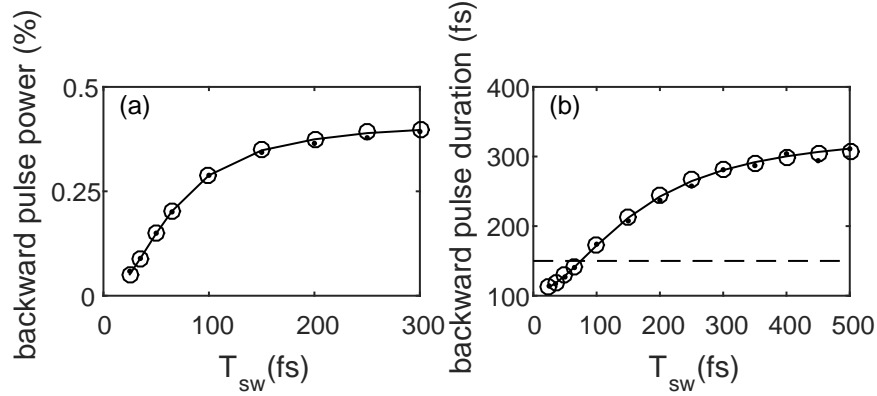}}
  \caption[]{(Color online) (a) Backward pulse (normalized) {\em peak} power and (b) duration for the perturbation~(\ref{eq:Delta_R_example1}). Here, $T_f = T_{pass} = 150$fs (horizontal dashed line) as a function of the switching time $T_{sw}$. Good agreement is found between the FDTD numerical simulations (dots), CMT simulations~(\ref{eq:CMT_t}) (circles) and the analytic solution~(\ref{eq:A_b}) (solid line) for a Gaussian-shaped pump pulse in time and space with a $n_{wg} = 1.5$, $2\mu$m wide single mode silica waveguide with $n_{eff} = 1.474$ at $\lambda_f = 2\mu$m, a pump wavelength of $1.5\mu$m
  and $\Delta n = 4 \cdot 10^{-3}$. } \label{fig:CMT_t_1}
\end{figure}

\begin{figure}[htbp]
  \centering{\includegraphics[scale=1]{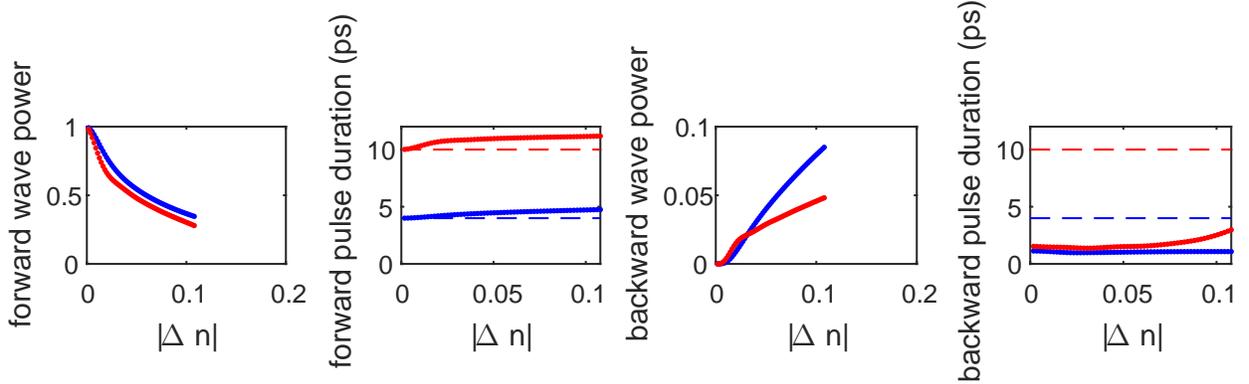}}
  \caption[]{(Color online) Results of numerical solution of the coupled mode equations~(\ref{eq:CMT_approx_1}) for the diffusive switching scheme~\cite{Sivan-COPS-switching-TBG}. (a) Backward pulse (normalized) {\em total} power and (b) duration as a function of $|\Delta n|$ for $n_{wg} = 2.5$, $T_{rise} = 300$fs, $T_{pass} = 150$fs, $T_{diff} = 600$fs; data shown for $T_f = 4$ps (blue) and $T_f = 10$ps (red). } \label{fig:diffusive_switching}
\end{figure}

\section{Discussion}\label{sec:discussion}
Our main result, Eq.~(\ref{eq:CMT_t}), is an extension of the standard (i.e., purely spatial) CMT to pulses and to spatio-temporal perturbations. Unlike previous attempts to derive such a model, our approach involves no approximation. In particular, it avoids the Slowly-Varying Envelope (SVE) approximation and the scalar mode assumption, and no restriction on the spatio-temporal profile is imposed. The effect of modal dispersion on mode evolution and on the coupling to other modes is fully taken into account. Thus, our approach can yield any required accuracy by retaining as many terms in the expansion as needed.

Compared with standard, purely spatial CMT, our approach has three additional features. First, the total (electric) field is written as the product of the modal amplitude and the modal profile in frequency domain (i.e., as a convolution of the corresponding time-domain expressions) rather than in the time domain, see Eq.~(\ref{eq:E_pulse}). This modification enables us to avoid the undesired effect of coupling between different pulsed wavepackets in the absence of a perturbation, as occurred in a few previous derivations\footnote{Such coupling is possible because modes of different order but different frequency are not necessarily orthogonal to each other. }, see e.g.,~\cite{temporal_CMT_Panoiu_FWMix}, or as occurs in some commercial FDTD software (see e.g.,~\cite{Lumerical}).

Second, material {\em and structural} dispersion are exactly taken into account via a series of derivatives of the propagation constant $\beta_m(\omega)$ and modal amplitude $A_m(t)$, which is identical to the one familiar from studies of the pulses in Kerr media~\cite{Agrawal-book,Fibich-book}. We note that in order to reach this familiar form, we partially lumped the effect of dispersion into the definition of the modal amplitude $A_m(t)$, see Eqs.~(\ref{eq:E_ansatz2}) and~(\ref{eq:A_m_hat_tilde}). In that regard, it should be noted that one can apply a Fourier transform over Eq.~(\ref{eq:CMT2}), i.e., over $\tilde{a}_m(z,\omega)$ rather than over $a_m(z,\omega)$, however, the resulting equation will not be standard - it will include the terms $\partial_{zz} \tilde{A}_m$, $\partial_z \tilde{A}_m$, $\partial_{zt} \tilde{A}_m$, $\partial_{ztt} \tilde{A}_m$ etc. and the definition of the modal amplitude in the time domain will be different from that of Eq.~(\ref{eq:E_ansatz2}). Indeed, one can retrieve the more familiar form by integrating over $z$, but then, the coupling will still differ from the familiar form. A comparison of the efficiency of these two formalism is beyond the scope of this paper.

Third, the coupling coefficient $\mathcal{K}_{m,n}$ takes a generalized form in which the modal profiles of the coupled modes are convolved with the spectral content of the perturbation. This effect accounts for the time derivatives of the ``refractive index change'' $\Delta R$ and of the dispersive nature of the perturbation may have a substantial effect only for rather short perturbations, strongly dispersive systems (e.g.,~\cite{temporal_CMT_Panoiu_FWMix,phc_wgs_slow_light,de_Rossi_pulse_compression_PhC_wgs,plasmonic_grating_slow_vg,MIM_EIT,plasmonic_slow_vg,plasmonic_slow_vg_Kuipers}), for perturbations with a complex spatial distribution and/or cases where the spatial overlap $\kappa(z,\omega,\omega')$ has a strongly asymmetric spectral dependence. It may also be significant in slow light regimes, where the correction to the group velocity may be substantial. While for optical frequencies these conditions may be difficult to achieve, in other spectral regimes, such as the microwave or the THz, it could be substantial. Indeed, to date, little work has been done on pulse propagation and time-varying media in the latter regime. Moreover, the generalized form of the coupling coefficient raises some open questions regarding the way to compute the coupling for modes $\vec{e}_m(\omega)$ which are multi-valued as, e.g., in the case of thin metallic circular wires or near cutoff or absorption lines where $\mathcal{K}_{m,n}$ may become complex or even purely imaginary.

Our derivation extends the work by the Bahabad group~\cite{temporal_CMT_Bahabad} by exactly accounting for the effect of structural dispersion, and allowing for more general perturbations, specifically, perturbations consisting of pulses rather than a discrete set of higher harmonics. Conveniently, our formalism requires only minimal modifications to that of~\cite{temporal_CMT_Bahabad}, but is more accurate and far more general. % Second, {\bf Panoiu ... removing artificial mode coupling,}
Our approach also extends the standard models of nonlinear wave propagation and mixing (e.g.,~\cite{Agrawal-book,Agrawal_Painter_review_OE_2007}) which typically involve only one mode in each frequency, to account for mode coupling. Indeed, our formulation would be especially useful for pulse propagation in nonlinear multimode fibers and waveguides~\cite{multimode_wg_experiments} where rich mode coupling occurs and complex spatio-temporal dynamics ensues. In that sense, our approach improves the accuracy of the dispersion calculation in~\cite{multimode_wg_theory-Kerr} and generalize it for additional nonlinear mechanisms.

We note that while standard CMT is completely analogous to quantum mechanical theory (by replacing $z$ with $t$), our extended perturbation theory does not have a quantum mechanical analogue. In that regard, our formalism allows the investigation of quantum phenomena such as coupling between discrete state and a continuum of states, Fano resonances etc. in a generalized context. These effects arise naturally for sufficiently short perturbations (in space and/or time), see e.g.,~\cite{Sivan-short-pulses_TBG}, where the bound modes can be coupled to the radiative modes, lying above the light line. These effects will be subject to future investigations.

Having stated the merits of our approach, it would be appropriate to recall its limitations. CMT becomes increasingly less effective as more modes are interacting in the system, in the presence of walk off between the interacting pulsed wave packets, or in cases of substantial frequency shifts (e.g.,~\cite{Cheskis_OL} for a Raman induced red-shift or~\cite{Travers_Russell} for a free-carrier induced blue shift). These effects are naturally taken into account within a self-consistent FDTD implementation. % (associated with separation of variables and long distance propagation) - what are the typical length scales for that? mm/cm probably.. ansatz not useful.. same would be for changes of dispersion due to frequency shifts.. so either correct dynamically, or better replace by an ansatz that does not involve a carrier wave - unidirectional .. maybe not even..}
A further complication of CMT arises for pulses of only a few cycles, in which case, the non-paraxiality terms have to be retained. In this case, one can instead employ the unidirectional flux formalism, so that the Slowly-Varying Envelope Approximation (SVEA) in time can be avoided while still keeping the formulation as a first order PDE, see~\cite{deep_gratings_PRE_96,Uni-directionality,Biancalana_1D_layered_media_n(t),Sivan-Pendry-article}. However, to date, the unidirectional formulation was implemented only for 1D systems~\cite{deep_gratings_PRE_96,Biancalana_1D_layered_media_n(t),Uni-directionality,Sivan-Pendry-letter,Sivan-Pendry-article}. In that sense, the most efficient and general formalism would be one that combined the CMT formulation described in this manuscript with unidirectional formulation. This too will be subject to future work.

In the final part of this section, it is necessary to emphasize the importance of the examples studied above. From the theoretical point of view, this is probably one of the first detailed studies of the interaction of several pulses of different durations and spatial extent. Indeed, the configuration studied is rather complex, as it involves 5 potentially different time scales (signal carrier period and duration, switching time, passage time, and the backward pulse time). In addition, the validation of the derived set of equations with exact FDTD is the first of its kind, at least to the best of our knowledge. Clearly, the more efficient implementation (with FC nonlinearity) is difficult to study with FDTD.

From the practical point of view, the configuration of the TBG provides a novel way to control the intensity, spatio-temporal shape and overall duration of a pulse of {\em arbitrary} central wavelength. This is potentially a more compact and flexible way compared with existing approaches for pulse shaping. In particular, since the backward pulse can be shorter than all other time scales in the system, this configuration provides a novel, efficient and simple way to temporally compress short pulses. Importantly, unlike spectral broadening based on self-phase modulation~\cite{Agrawal-book}, the spectral broadening resulting from the spectral exchange in the current case, leads to temporal compression without the need for further propagation in a dispersive medium. It is also cleaner, hence more effective, in the sense that the generated backward pulses are nearly transform-limited.

More generally, the same formulation and results are expected for different configurations, involving, for example, long Bragg gratings for broadband mode conversion~\cite{Fallnich-Hellwig}.

% {\bf COPS - switching pulses were of wide spatial extent? }

\section*{Acknowledgements}
The authors would like to acknowledge the critical contribution of S. Pinhas to the initial derivations and many useful discussions with A. Ishaaya.

%{\bf  and A. Bahabad.
%
%additions after review - add Alon, refer to Horak + "PRL", update Fink + Armenians, Timor?
%
%other interested: Biancalana (or maybe wait for Marat's paper?), Panoiu, Ulf, Tom Philbin, Kinsler, Yoni Jaffa, Longhi, Menyuk, Fan, Boyd, Fink, Yariv, Yeh, Aceves, Agrawal, Painter, de Sterke, Sipe, Joachim.
%
%\bigskip
%
%seek opinion of Biancalana, Horsely/Philbin, Peter Halevi. Fan, Notomi, Breuck + all the people Alon quotes.. Winful.
%
%\bigskip
%
%When done, send to turks, Kinsler?, Mosk/Vos/Euser/Georgios, D. Pile, Agrawal/Xiao, Painter. Lipson's people. Gauthier? Ulf + Tom, Horak + Poletti, Mafi, Wise, Christodoulides, Wabnitz, ...
%
%fix factor 2; 2 additional points shai has to check. Armen's comments and refs.
%
%}

\appendix

\section{Derivation of Eq.~(\ref{eq:CMT_t})}\label{app:FT}
In order to derive an equation for the ``mode amplitude'' in the time domain, $A_m(t)$, and separate different orders of dispersion, we treat the two sides of Eq.~(\ref{eq:CMT2}) separately.

Using the chain rule and definition~(\ref{eq:A_m_hat_tilde}), the terms on the LHS of Eq.~(\ref{eq:CMT2}) can be rewritten as
\begin{eqnarray}
&& e^{i \left(\beta_m - \beta_{m,0}\right) z} \frac{\partial^2}{\partial z^2} \tilde{a}_m(z,\omega - \omega_0) = \nn \\
&& \frac{\partial^2}{\partial z^2} a_m(z,\omega - \omega_0) - 2 i \left(\beta_m - \beta_{m,0}\right) e^{i \left(\beta_m - \beta_{m,0}\right) z} \frac{\partial}{\partial z} \tilde{a}_m(z,\omega - \omega_0) + \left(\beta_m - \beta_{m,0}\right)^2 a_m(z,\omega - \omega_0), \nn
\end{eqnarray}
and
\begin{eqnarray}
2 i \beta_m e^{i \left(\beta_m - \beta_{m,0}\right) z} \frac{\partial}{\partial z} \tilde{a}_m(z,\omega - \omega_0) = 2 i \beta_m \frac{\partial}{\partial z} a_m(z,\omega - \omega_0) + 2 \beta_m \left(\beta_m - \beta_{m,0}\right) a_m(z,\omega - \omega_0). \nn
\end{eqnarray}
Combining the above shows that the LHS of Eq.~(\ref{eq:CMT2}) becomes
\begin{eqnarray}\label{eq:CMT_RHS_standard}
&=& \frac{\partial^2}{\partial z^2} a_m(z,\omega - \omega_0) + 2 i \beta_{m,0} e^{i \left(\beta_m - \beta_{m,0}\right) z} \frac{\partial}{\partial z} \tilde{a}_m + \left(\beta_m - \beta_{m,0}\right)^2 a_m \nn \\
&=& ... \ = \ \frac{\partial^2}{\partial z^2} a_m + 2 i \beta_{m,0} \frac{\partial}{\partial z} a_m + \left(\beta_m^2 - \beta_{m,0}^2\right) a_m.
\end{eqnarray}
We now apply a Fourier transform $\mathcal{F}_\omega^t$ over Eq.~(\ref{eq:CMT_RHS_standard}) and get
\begin{eqnarray}\label{eq:LHS_t}
\!\!\!\!\! \!\!\!\!\!  \!\!\!\!\! e^{- i \omega_0 t} \left[\frac{\partial^2}{\partial z^2} A_m(z,t) + 2 i \beta_{m,0} \frac{\partial}{\partial z} A_m + i \frac{2\beta_{m,0}}{v_{g,m}} \frac{\partial}{\partial t} A_m - \left(\frac{1}{v_{g.m}^2} + \beta_{m,0} \beta''_{m,0}\right) \frac{\partial^2}{\partial t^2} A_m + \sum_{q=3}^\infty \alpha_{q,m} \left(i\frac{\partial}{\partial t}\right)^q A_m\right], \nn \\
\end{eqnarray}
where $\alpha_{q,m} \equiv \frac{1}{q!}\frac{d^q (\beta_m^2)}{d \omega^q}\big|_{\omega_0}$ are the coefficients of the high-order dispersion terms. Note that Eq.~(\ref{eq:LHS_t}) is now identical to the result obtained in the {\em exact} derivation of dispersive terms for free space propagation, as derived in~\cite[Eq.~(35.14), p. 746]{Fibich-book}~\footnote{Indeed, it can be shown that the standard derivation, as e.g., described in~\cite{Agrawal-book}, is correct only because the error introduced by the approximation $\beta^2 - \beta_0^2$ is cancelled by a premature neglect of the non-paraxiality term, see~\cite{Fibich-dispersion}. }. Then, when one performs the transformation to the moving frame and neglects the non-paraxiality and spatio-temporal coupling terms~\cite{Jain-Tzoar}, the (more) familiar (but approximate) dispersive equation of~\cite{Agrawal-book} is obtained.

On the RHS of Eq.~(\ref{eq:CMT2}), we expand all the slowly-varying functions of $\omega$, namely, $\kappa_{m,n}$ (i.e., avoiding the explicit expansion of $\vec{e}_m$), $\beta_m$, and $\omega$ itself in a Taylor series near $\omega_0$, while the rapidly-varying functions $\tilde{a}_n$ % \footnote{$e^{i \beta_{m,0} z}$ is just constant.. }
are left as they are~\footnote{Note that the slowly-varying exponential factor $e^{i (\beta_m(\omega) - \beta_{m,0}) z}$ is lumped into $\tilde{a}_n$ so that it is not expanded separately. }. Specifically, the terms on the RHS of Eq.~(\ref{eq:CMT2}) become
\begin{eqnarray}\label{eq:Taylor_omega2}
\omega |\beta_m(\omega)| = \omega_0 |\beta_{m,0}| + (\omega - \omega_0)\left(|\beta_{m,0}| + \omega_0\bigg|\frac{\partial \beta_m(\omega)}{\partial \omega}\bigg|_{\omega = \omega_0}\right) + O\left((\omega - \omega_0)^2\right), \nn
\end{eqnarray}
and
\begin{eqnarray}\label{eq:Taylor_C}
&& c_{m,n}(z,\omega,\omega - \omega_0) = \int d\omega' \kappa_{m,n}(z,\omega_0,\omega') e^{i \beta_n(\omega - \omega') z} \tilde{a}_n(z,\omega - \omega' - \omega_0) \nn \\
&& \quad \quad + (\omega - \omega_0) \int d\omega' \left[e^{i \beta_n(\omega - \omega') z} \tilde{a}_n(z,\omega - \omega' - \omega_0)\right] \frac{\partial}{\partial \omega}\left(\kappa_{m,n}(z,\omega,\omega')\right)\bigg|_{\omega_0} + O\left((\omega - \omega_0)^2\right), \nn
\end{eqnarray}
where
\begin{eqnarray}\label{eq:kappa_app}
\kappa_{m,n}(z,\omega,\omega') &=& \int d\vec{r}_\perp \Delta \epsilon(\vec{r}_\perp,z,\omega') \left[\vec{e}_m^*(\vec{r}_\perp,\omega) \cdot \vec{e}_n(\vec{r}_\perp,\omega - \omega')\right].
\end{eqnarray}
Thus, overall, the RHS of Eq.~(\ref{eq:CMT2}) can be rewritten as
\begin{eqnarray}\label{eq:CMT_app}
- \frac{e^{- i \beta_{m,0} z}}{2} \sum_n d_{m,n}^{(0)}(\omega) + (\omega - \omega_0) d_{m,n}^{(1)}(\omega) + ...,
\end{eqnarray}
where
\begin{eqnarray}
d_{m,n}^{(0)}(\omega) &=& \omega_0 |\beta_{m,0}| c_{m,n}(z,\omega_0,\omega - \omega_0) = \omega_0 |\beta_{m,0}| \int d\omega' \kappa_{m,n}(z,\omega_0,\omega  -\bar{\omega}) \ e^{i \beta_n(\bar{\omega}) z} \tilde{a}_n(z,\bar{\omega} - \omega_0), \nn \\ \\
d_{m,n}^{(1)}(\omega) &=& \left(|\beta_{m,0}| + \omega_0\bigg|\frac{\partial \beta_m(\omega)}{\partial \omega}\bigg|_{\omega = \omega_0}\right) \int d\omega' \kappa_{m,n}(z,\omega_0,\omega') \left[e^{i \beta_n(\omega - \omega') z} \tilde{a}_n(z,\omega - \omega' - \omega_0)\right] \nn \\
&+& \omega_0 |\beta_{m,0}| \int d\omega'  \frac{\partial}{\partial \omega}\left(\kappa_{m,n}(z,\omega,\omega')\right)\bigg|_{\omega_0} \left[e^{i \beta_n(\omega - \omega') z} \tilde{a}_n(z,\omega - \omega' - \omega_0)\right],
\end{eqnarray}
and similarly for the next order terms. Fourier transforming Eq.~(\ref{eq:LHS_t}) gives
\begin{eqnarray}\label{eq:RHS_t_app}
- \frac{e^{- i \beta_{m,0} z}}{2} \sum_j D_{m,n}^{(j)}(t),
\end{eqnarray}
where
\begin{eqnarray}
D^{(0)}_{m,n}(t) &=& \omega_0 |\beta_{m,0}| e^{- i \omega_0 t} \int d\omega e^{- i (\omega - \omega_0) t} \int d\omega' \kappa_{m,n}(z,\omega_0,\omega') \left[e^{i \beta_n(\omega - \omega') z} \tilde{a}_n(z,\omega - \omega' - \omega_0)\right] \nn \\
&=& \omega_0 |\beta_{m,0}| e^{- i \omega_0 t + i \beta_{n,0} z} \int d\omega' \kappa_{m,n}(z,\omega_0,\omega') e^{- i \omega' t} \int d\bar{\omega} e^{i \left[\beta_n(\bar{\omega}) - \beta_{n,0}\right]z} \tilde{a}_n(z,\bar{\omega} - \omega_0) e^{- i (\bar{\omega} - \omega_0) t} \nn \\
&=& \omega_0 |\beta_{m,0}| \mathcal{K}_{m,n}(z,\omega_0,t) e^{- i \omega_0 t + i \beta_{n,0} z} A_n(z,t),
\end{eqnarray}
and where
\begin{eqnarray}\label{eq:kappa_t_app}
\mathcal{K}_{m,n}(z,\omega_0,t) &=& \mathcal{F}_{\omega'}^t [\kappa_{m,n}(z,\omega_0,\omega')] \equiv \int d\omega' \kappa_{m,n}(z,\omega_0,\omega') e^{- i \omega' t} \nn \\
&=& \int d\omega' \int d\vec{r}_\perp \left[\vec{e}_m^*(\vec{r}_\perp,\omega_0) \cdot \vec{e}_n(\vec{r}_\perp,\omega_0 - \omega')\right] \Delta \epsilon(\vec{r}_\perp,z,\omega')e^{- i \omega' t}.
\end{eqnarray}
Similarly,
\begin{eqnarray}
\!\!\!\! \!\!\!\! \!\!\!\! \!\!\!\! \!\!\!\! D^{(1)}_{m,n}(t) &=& \left(|\beta_{m,0}| + \omega_0\bigg|\frac{\partial \beta_m(\omega)}{\partial \omega}\bigg|_{\omega_0}\right) e^{- i \omega_0 t} \int d\omega (\omega - \omega_0) e^{- i (\omega - \omega_0) t} \int d\omega' \kappa_{m,n}(z,\omega_0,\omega') \left[e^{i \beta_n(\omega - \omega') z} \tilde{a}_n(z,\omega - \omega' - \omega_0)\right] \nn \\
&+& \omega_0 |\beta_{m,0}| e^{- i \omega_0 t} \int d\omega e^{- i (\omega - \omega_0) t} (\omega - \omega_0) \int d\omega' \frac{\partial}{\partial \omega}\left(\kappa_{m,n}(z,\omega,\omega')\right)\bigg|_{\omega_0} \left[e^{i \beta_n(\omega - \omega') z} \tilde{a}_n(z,\omega - \omega' - \omega_0)\right] \nn \\
&=& ... \ = \ i \left[\left(|\beta_{m,0}| + \omega_0\bigg|\frac{\partial \beta_m(\omega)}{\partial \omega}\bigg|_{\omega_0}\right)\mathcal{K}_{m,n}(z,\omega_0,t) + \omega_0 |\beta_{m,0}| \frac{\partial}{\partial \omega}\left(\mathcal{K}_{m,n}(z,\omega,t)\right)\bigg|_{\omega_0}\right] e^{- i \omega_0 t  + i \beta_{n,0} z} \frac{\partial}{\partial t} A_n(z,t).
\end{eqnarray}
The next order terms can be calculated the same way.

\section{Derivation of coupled mode equations for the generalized Helmholtz equation}\label{app:vector_Helmholtz}
{\color{black} In the most general case, the Maxwell equations are equivalent to the vector wave equation
\begin{equation}\label{eq:vector_waeq_app}
\left[\nabla^2 - \nabla (\nabla \cdot )\right]\vec{E}(\vec{r}_\perp,z,t) = \mu \frac{\partial^2}{\partial t^2} \vec{D}(\vec{r}_\perp,z,t).
\end{equation}
Substituting Eq.~(\ref{eq:E_ansatz2}) in Eq.~(\ref{eq:vector_waeq_app}) yields, for the LHS,
\begin{eqnarray}\label{eq:E_m_vec_LHS}
&& \left(\left[\nabla^2_\perp + \frac{\partial^2}{\partial z^2}\right] - \nabla(\nabla \cdot)\right) \left(\sum_m \int_{-\infty}^\infty d\omega e^{- i \omega t + i \beta_m(\omega) z} \tilde{a}_m(z,\omega - \omega_0) \vec{e}_m(\vec{r}_\perp,\omega)\right) = \nn \\
&& \sum_m \int_{-\infty}^\infty d\omega e^{- i \omega t} \left(\left[\nabla^2_\perp + \frac{\partial^2}{\partial z^2}\right] \left[e^{i \beta_m(\omega) z} \tilde{a}_m \vec{e}_m\right] - \nabla(\nabla \cdot) \left[ e^{i \beta_m z} \tilde{a}_m \vec{e}_m\right]\right), \nn
\end{eqnarray}
and for the RHS,
\begin{equation}\label{eq:dp_app}
- \mu \int_{-\infty}^\infty d\omega \omega^2 e^{- i \omega t} \sum_n \int_{-\infty}^\infty d\omega' \left[\epsilon(\vec{r}_\perp,\omega') + \Delta\epsilon(\vec{r}_\perp,z,\omega')\right] \vec{e}_n(\vec{r}_\perp,\omega - \omega') e^{i \beta_n(\omega - \omega') z} \tilde{a}_n(z,\omega - \omega' - \omega_0). \nn
\end{equation}
The integrand of the laplacian terms reads
\begin{eqnarray}\label{eq:E_m_vec_LHS_laplace}
&& \left[\nabla^2_\perp + \frac{\partial^2}{\partial z^2}\right] \left[\tilde{a}_m(z,\omega - \omega_0) e^{i \beta_m z} \vec{e}_m(\vec{r}_\perp,\omega)\right] = \left[\tilde{a}_m(z,\omega - \omega_0) e^{i \beta_m z}\right] \nabla^2_\perp  \vec{e}_m(\vec{r}_\perp,\omega) \nn \\
&+& \vec{e}_m(\vec{r}_\perp,\omega) \left[\frac{\partial^2}{\partial z^2} \tilde{a}_m(z,\omega - \omega_0) + 2 i \beta_m \frac{\partial}{\partial z} \tilde{a}_m(z,\omega - \omega_0) - \beta_m^2 \tilde{a}_m\right] e^{i \beta_m z},
\end{eqnarray}
while the integrand of the grad-div terms is
\begin{eqnarray}\label{eq:grad-div}
&=& \nabla \left[e^{i \beta_m z} \tilde{a}_m(z,\omega - \omega_0) \nabla_\perp \left(\hat{r}_\perp \cdot \vec{e}_m(\vec{r}_\perp,\omega)\right) + \left(\hat{z} \cdot \vec{e}_m(\vec{r}_\perp,\omega)\right) e^{i \beta_m z}(i \beta_m \tilde{a}_m + \frac{\partial}{\partial z} \tilde{a}_m)\right] \nn \\
&=& ... = e^{i \beta_m z} \tilde{a}_m(z,\omega - \omega_0) (\nabla_\perp \left(\hat{r}_\perp \cdot \vec{e}_m(\vec{r}_\perp,\omega)\right) + i \beta_m \left[\nabla_\perp (\hat{z} \cdot \vec{e}_m(\vec{r}_\perp,\omega)) \hat{r}_\perp + \nabla_\perp \left(\hat{r}_\perp \cdot \vec{e}_m(\vec{r}_\perp,\omega)\right)\hat{z} \right] \nn \\
&-& \beta_m^2 \tilde{a}_m \left(\hat{z} \cdot \vec{e}_m(\vec{r}_\perp,\omega)\right) \hat{z} \nn \\
&+& e^{i \beta_m z} \frac{\partial}{\partial z} \tilde{a}_m \left[\nabla_\perp (\hat{z} \cdot \vec{e}_m(\vec{r}_\perp,\omega)) \hat{r}_\perp + \nabla_\perp \left(\hat{r}_\perp \cdot \vec{e}_m(\vec{r}_\perp,\omega)\right)\hat{z} + 2 i \beta_m (\hat{z} \cdot \vec{e}_m(\vec{r}_\perp,\omega))\hat{z} \right] \nn \\
&+& \left(\hat{z} \cdot \vec{e}_m(\vec{r}_\perp,\omega)\right) \frac{\partial^2}{\partial z^2} \tilde{a}_m \hat{z},
\end{eqnarray}
where we used $\nabla = \nabla_\perp + \frac{\partial}{\partial z}$.

All the terms proportional to $\tilde{a}_m$ vanish as they constitute the equation satisfied by the mode profile $\vec{e}_m$. The remaining terms from the grad-div term~(\ref{eq:grad-div}) add to the remaining $z$ derivatives in Eq.~(\ref{eq:E_m_vec_LHS_laplace}), and somewhat change the transverse function compared with the case~(\ref{eq:CMT1}) in which the grad-div term vanishes identically. As a result, the normalization of the modes should be defined slightly differently, and the various terms in the resulting CMT formulation attain somewhat different weights compared with the regular case~(\ref{eq:CMT_t}). An alternative derivation, which is more customary is based on the transverse fields only, see~\cite{Marcuse-book,vectorial_CMT}.

\section{The failure of the averaged wavenumber ansatz}~\label{app:standard_derivations}
Instead of using the correct ansatz~(\ref{eq:E_ansatz1})-(\ref{eq:E_ansatz2}), we now adopt a somewhat simpler ansatz, in which the wavenumber of the various frequency components within every wave packet $m$ is set to the central value, i.e.,
\begin{equation}\label{eq:standard_ansatz}
\vec{E}(\vec{r}_\perp,z,t) = \sum_m e^{- i \omega_0 t + i \beta_{m,0} z} B_m(z,t) \vec{e}_m(\vec{r}_\perp,\omega_0).
\end{equation}
Substituting in the Helmholtz equation yields
\begin{eqnarray}\label{eq:scalar_waeq_app}
\left(\nabla_\perp^2 + \frac{\partial^2}{\partial z^2}\right) &\sum_m& B_m(z,t) e^{- i \omega_0 t + i \beta_{m,0}z} \vec{e}_m(\vec{r}_\perp,\omega_0) = \mu \frac{\partial^2}{\partial t^2} P(\vec{r}_\perp,z,t) \nn \\
&=& - \mu \int_{-\infty}^\infty d\omega e^{- i \omega t} \omega^2 \epsilon(\vec{r}_\perp,\omega) \mathcal{F}_t^\omega \left[\sum_n B_n(z,t) e^{- i \omega_0 t + i \beta_{n,0}z} \vec{e}_n(\vec{r}_\perp,\omega_0) \right] \nn \\
&=& - \mu \sum_n e^{i \beta_{n,0}z} \vec{e}_n(\vec{r}_\perp,\omega_0) \int_{-\infty}^\infty d\omega e^{- i \omega t} \omega^2 \epsilon(\vec{r}_\perp,\omega) b_n(z,\omega - \omega_0),
\end{eqnarray}
where $b_m(\omega) = \mathcal{F}^\omega_t(B_m(t))$ and where $P(\vec{r}_\perp,t)$ is given by Eq.~(\ref{eq:P}); here we assume the system is unperturbed. Fourier transforming Eq.~(\ref{eq:scalar_waeq_app}) and performing the differentiations gives
\begin{eqnarray}\label{eq:scalar_waeq_app2}
\sum_m e^{i \beta_{m,0} z} \vec{e}_m \frac{\partial^2}{\partial z^2} b_m &+& 2 i \beta_{m,0} e^{i \beta_{m,0} z} \vec{e}_m \frac{\partial}{\partial z} b_m - \beta_{m,0}^2 b_m e^{i \beta_{m,0} z} \vec{e}_m + b_m e^{i \beta_{m,0} z} \nabla_\perp^2 \vec{e}_m \nn \\
&+& \mu \omega^2 \epsilon(\vec{r}_\perp,\omega) \sum_n b_n(z,\omega - \omega_0) e^{i \beta_{n,0}z} \vec{e}_n(\vec{r}_\perp,\omega_0) = 0.
\end{eqnarray}
Importantly, as in the standard derivation for uniform nonlinear media~\cite{Fibich-book} or as in the derivation for single mode nonlinear fibers~\cite{Agrawal-book}, for $n = m$, the last three terms on the LHS of Eq.~(\ref{eq:scalar_waeq_app2}) cancel {\em partially}, leaving on the LHS of Eq.~(\ref{eq:scalar_waeq_app2}) the term
\begin{equation}
\left(\mu \epsilon(\vec{r}_\perp,\omega) \omega^2 - \beta_{m,0}^2\right) b_m e^{i \beta_{m,0} z} \vec{e}_m,
\end{equation}
as well as terms proportional to $\vec{e}_{n \ne m}$. In the next step, the dependence on the transverse coordinates is removed by multiplying the remaining terms by $\vec{e}_m^*$. The $n = m$ term is expanded in a Taylor series, and yields the well-known dispersion terms. However, due to the dependence on $\vec{r}_\perp$, terms corresponding to different ($m \ne n$) modes {\em do not vanish}. Thus, {\em the averaged wavenumber ansatz~(\ref{eq:standard_ansatz}) gives rise to mode coupling even in the absence of perturbation}. Indeed, such terms were observed in~\cite{temporal_CMT_Panoiu_FWMix}, and then averaged out by hand, and neglected altogether in~\cite{temporal_CMT_Bahabad}. The reason for the appearance of these coupling terms is that the factor $e^{i \beta_{m,0} z}$ in the ansatz~(\ref{eq:standard_ansatz}) introduces phase errors to all frequency components within the wave packet $m$ (except for the central component). Thus, while the averaged wavenumber based ansatz~(\ref{eq:standard_ansatz}) is appropriate in the absence of (material and structural) dispersion, it is clearly inappropriate in the presence of dispersion, thus, is a fundamental deficiency of the standard approaches. As shown in Section~\ref{sec:derivation}, this undesired effect can be avoided altogether by adopting a more careful ansatz~(\ref{eq:E_ansatz1}).

The length scale of this somewhat ``artificial`` coupling is proportional to the index contrast in the waveguide, hence, it may be difficult to observe in numerical simulations performed on short waveguide segments. In any case, we believe that no exact numerical simulations validating previous coupled-mode models for pulses was made before, so that the inaccuracies incurred by this undesired artificial mode coupling were not observed before. There are observed, however, in many standard numerical software based on FDTD, which employ the averaged wavenumber ansatz, see e.g.,~\cite{Lumerical}.

} % end of blue section

%\section{Points to discuss with Alon}
%\begin{itemize}
%%   \item realize subtlety/accuracy of their ansatz? no...
%%   \item their derivation applied mostly for wg (or TE/TM/2D) geometries.. - did they mean to? mostly they deal with large systems..
%%   \item did their input and perturbation have some spectral width? usually not
%%   \item is our reference to their formulation correct? yes.
%%   \item did they not note the cross-coupling? no..
%%   \item can our derivation solve any problem for them? yes, the plasmonic cases!
%%   \item they do not calculate even vg self-consistently.. he did not realize..
%  \item Hardy - consulted him?
%\end{itemize}

% \bibliography{D:/MyDocs/Research/my_bib}
% laptop
% \bibliography{C:/Users/Yonatan/Documents/Research/Ultrafast_stuff/Time_reversal/Israel_2013/diffusion_write_erase_manuscript/my_bib}
% \bibliography{C:/Users/Yonatan/Documents/Research/my_bib}
% \bibliographystyle{unsrt}

\end{document}